\documentclass[MSNbibl,seceqn,nameyear,dvips]{arxstspdf}
\usepackage{flushend}
\usepackage{stfloats}
\usepackage{graphicx}

\volume{27}
\issue{1}
\pubyear{2012}
\firstpage{95}
\lastpage{114}
\doi{10.1214/11-STS374}

\makeatletter

\def\bsuffix #1{#1}

\newtheorem{theorem}{Theorem}
\newproclaim{remark}{Remark}

\newcommand{\beq}{\begin{equation}}
\newcommand{\eeq}{\end{equation}}

\newcommand{\bdzero}{\mathbf{0}}

\newcommand{\bA}{\mathbf{A}}
\newcommand{\bB}{\mathbf{B}}
\newcommand{\bC}{\mathbf{C}}
\newcommand{\bD}{\mathbf{D}}
\newcommand{\be}{\mathbf{e}}
\newcommand{\bG}{\mathbf{G}}
\newcommand{\bI}{\mathbf{I}}
\newcommand{\bK}{\mathbf{K}}
\newcommand{\bP}{\mathbf{P}}
\newcommand{\bT}{\mathbf{T}}
\newcommand{\bu}{\mathbf{u}}
\newcommand{\bV}{\mathbf{V}}
\newcommand{\bX}{\mathbf{X}}
\newcommand{\bY}{\mathbf{Y}}
\newcommand{\bZ}{\mathbf{Z}}

\newcommand{\bSi}{\bolds{\Sigma}}
\newcommand{\bth}{\bolds{\theta}}

\newcommand{\bla}{\bolds{\lambda}}
\newcommand{\bPsi}{\bolds{\Psi}}
\newcommand{\bbe}{\bolds{\beta}}
\makeatother

\begin{document}
\begin{frontmatter}
\vspace*{6pt}
\title{Small Area Shrinkage Estimation}
\runtitle{Small Area Estimation}

\begin{aug}
\author[a]{\fnms{G.} \snm{Datta}\ead[label=e1]{gauri@uga.edu}}
\and
\author[b]{\fnms{M.} \snm{Ghosh}\corref{}\ead[label=e2]{ghoshm@stat.ufl.edu}}
\runauthor{G. Datta and M. Ghosh}

\affiliation{University of Georgia and University of Florida}

\address[a]{G. Datta is Professor, Department of Statistics, University of Georgia, Athens,
Georgia 30602-7952, USA \printead{e1}.}
\address[b]{M. Ghosh is Distinguished Professor,
Department of Statistics, University of Florida, Gainesville, Florida 32611-8545,
USA \printead{e2}.}

\end{aug}

%
\begin{abstract}
The need for small area estimates is increasingly felt in both the
public and
private sectors in order to formulate their strategic plans. It is now widely
recognized that direct small area survey estimates are highly
unreliable owing
to large standard errors and coefficients of variation. The reason behind
this is that a survey is usually designed to achieve a specified level of
accuracy at a higher level of geography than that of small areas. Lack of
additional resources makes it almost imperative to use the same data to
produce small area estimates. For example, if a survey is designed to estimate
per capita income for a~state, the same survey data need to be used to produce
similar estimates for counties, subcounties and census divisions within that
state. Thus, by necessity, small area estimation needs explicit, or at least
implicit, use of models to link these areas. Improved small area
estimates are
found by ``borrowing strength'' from similar neighboring areas.

The key to small area estimation is shrinkage of direct estimates toward
some regression estimates obtained by using in addition administrative
records and other available sources of information. These shrinkage
estimates can often be motivated from both a Bayesian and a frequentist
point of view, and indeed in this particular context, it is possible to obtain
at least an operational synthesis between the two paradigms. Thus, on
one hand,
while small area estimates can be developed using a hierarchical
Bayesian or an
empirical Bayesian approach, similar estimates are also found using the theory
of best linear unbiased prediction (BLUP) or empirical best linear unbiased
prediction (EBLUP).

The present article discusses primarily normal theory-based small area
estimation techniques, and attempts a synthesis between both the Bayesian
and the frequentist points of view. The results are mostly discussed for
random effects models and their hierarchical Bayesian counterparts.
A few miscellaneous remarks are made at the end describing the current
research for more complex models including some nonnormal ones. Also provided
are some pointers for future research.
\end{abstract}

%

\end{frontmatter}

\section{Introduction}\label{sec1}\vspace*{-5pt}

Small area estimation has become a topic of growing importance in
recent years. The need for such estimates is increasingly felt in both
the public and private sectors in order to formulate their strategic
plans. For instance, to address emerging or existing social issues,
many national governments have passed laws that require production of
reliable and up-to-date small area estimates on a regular basis. As an
example, in the early 1990s, the U.S. Congress passed a law requiring
the Secretary of Commerce to produce and publish, at least every two
years, starting in 1996, current small area estimates related to the
incidence of poverty for states, counties, local jurisdictions of
governments and school districts. In the private sector, businesses,
especially the smaller ones, make decisions based on local income,
population and environmental data to evaluate markets for new products
and to determine areas for the location, expansion and contraction of
their activities.

Small areas may refer to small geographical \mbox{areas} such as counties,
subcounties, census tracts, etc. Alternately, they may also refer to
small domains cross-classified by age, sex and other demographic
characteristics. Other than ``small areas'' and ``small domains,''
often the terms ``local areas,'' ``subdomains'' and ``substates''
are used interchangeably. Throughout this article, we will use the
term ``small area,'' possibly the most popular usage of the term,
especially in survey sampling.

Shrinkage estimators have even a longer history than small area
estimators. An exact definition of these estimators is hard to come by.
Lemmer (\citeyear{le88}) in his \textit{Encyclopedia of Statistical Sciences}
article characterized shrinkage estimators as ones obtained through
modification of some standard estimators, for example, maximum
likelihood estimator
(MLE), uniformly minimum variance unbiased estimator\break (UMVUE),
least squares estimator, etc., in order to minimize some desirable
criterion such as mean squared error (MSE), quadratic risk, bias, etc.
With these objectives in mind, shrinkage estimators can be interpreted
in a very broad sense. In particular, the best linear unbiased predictors
(BLUP's), empirical best linear unbiased predictors (EBLUP's), empirical
Bayes (EB), hierarchical Bayes (HB), and possibly a host of other
estimators fall within this general category. One common feature of all
these estimators is that they are usually weighted averages of one of
the aforementioned standard estimators and some other estimator reasonable\vadjust{\goodbreak}
under an appropriate model. Weights to these estimators are determined
with the
objective of meeting some ``optimality'' criterion.

Shrinkage estimates have a natural place in small area estimation where
direct estimates such as the MLE, UMVUE, etc., are usually unreliable
owing to
large standard errors and coefficients of variation associated
with them. The reason behind this is that the original survey was targeted
to achieve accuracy at a higher order of aggregation than that of
small areas. Due to limited resources, the same survey data need
to be used for producing small area estimates. This necessitates
``borrowing strength'' from similar other small areas with the
objective of ``increasing the effective sample size'' in order to obtain
estimates of increased precision.

The early small area estimators achieved this objective by shrinking
the area-specific direct estimators (e.g., county-specific
averages) toward some overall estimator (e.g., the state average).
Later, with the availability of auxiliary information from administrative
records and other sources, the direct estimators are now usually shrunk
toward some estimated regression surface. This shrinking process needs explicit
(or at least implicit) use of models.

Bayesian estimators have been in existence for more than two centuries.
Very often, they can be regarded as shrinkage estimators, shrinking, for
example, the sample mean toward the prior mean. The BLUP and EBLUP
estimators developed by Henderson (\citeyear{Hen53}) for mixed linear models
are also genuine shrinkage estimators, shrinking the direct estimators
toward some regression estimators. However, as the title of this special
issue suggests, the name ``shrinkage'' possibly was coined with the
seminal paper of Stein (\citeyear{Ste56}). Stein introduced shrinkage
estimators
to estimate a multivariate normal mean vector and proved under the sum of
squared error loss their domination over the sample mean vector in
three or
higher dimensions. He gave a purely decision-theoretic motivation
of his result, and was implicitly considering a balanced
one-way ANOVA model for random effects. The original result of Stein
involved shrinking the sample mean toward some guessed value of
the population mean. Later extensions of his ideas
due to Lindley (\citeyear{Lin}) and Stein (\citeyear{Ste62}) led to
shrinkage toward an overall
average, and more generally to a regression surface, still with
balanced data.
Stein's estimators gained immense popularity in the 1970s when Efron
and Morris,
in a series of articles, gave interesting EB\vadjust{\goodbreak}
interpretation of these estimators (see, e.g., Efron and Morris,
\citeyear{EfrMor73}).
A pioneering extension of Stein's ideas in the small area estimation
context is
due to Fay and Herriot (\citeyear{FayHer79}) in their highly referred
article. The
paper showed how Stein-type results (without necessarily the \textit{exact}
dominance consideration) could be extended to unbalanced random
effect regression models with tremendous potential for
application.\looseness=1

It is near impossible to cover all aspects of small area estimation in a
single review article. Our primary focus will be on one-way
random effects regression models, and connecting the ideas of BLUP and
EBLUP with HB and EB estimators.
These models are usually referred to in the small area literature as
``area-level'' models where one begins with some small area summary
statistics, and
tries to improve on these estimators by shrinking them toward some regression
surface. This is in contrast to the so-called ``unit-level'' models where
one has data available for the sampled units within a small area. We will
barely touch upon the latter. Another component of research
which has received scant attention in the small area
literature is the development of EB confidence intervals. We will
discuss this topic also at some length. For a detailed exposure to
small area
estimation, the reader is referred to the recent book of Rao (\citeyear
{Rao03N1})
and the
review articles of Ghosh and Rao (\citeyear{GhoRao94}), Pfeffermann
(\citeyear{Pfe02}), Rao
(\citeyear{Rao99,Rao03N2})
and Datta (\citeyear{Dat}).

The outline of the remaining sections is as follows. In Section \ref
{sec2}, we discuss
balanced one-way random effects regression models, and discuss the connection
between the BLUP's, EBLUP's, HB, EB, and in particular, the Stein-type
shrinkage estimators. Section~\ref{sec3} extends these results to
unbalanced one-way
models, and compares and contrasts both HB and EB estimators in this setup.
MSE approximation of small area estimators is also discussed in this section.
Section \ref{sec4} discusses multivariate small area shrinkage
estimators, and discusses
one particular application related to adjustment of census counts.
Section \ref{sec5}
discusses EB confidence intervals for both balanced and unbalanced data.
Section \ref{sec6} gives a brief account of unit-level models for small
area estimation.
Section \ref{sec7} contains a few other small area models such as
measurement error
models and generalized linear models. This section contains also a discussion
of balanced loss functions in the context of small area estimation.
Section \ref{sec8} contains a summary of the results presented, and
provides a few
pointers toward topics for future research.

\section{Shrinkage Estimators for~Balanced Data}\label{sec2}\vspace*{-3pt}
The primary objective of this section is to introduce shrinkage
estimators of
small area means
under different paradigms, and point out the interrelationship between them.
The corresponding uncertainty measures are also compared. We
begin with the following model.

Let $y_i$ $(i=1,\ldots, m)$ denote the area-level survey estimators for
the $m$ small areas.
Consider the model
%
%
\begin{eqnarray}\label{e2p1}
y_i|\theta_i &\stackrel{\mathit{ind}}\sim& N(\theta_i, V),\quad
\mbox{and}
\nonumber
\\[-8pt]
\\[-8pt]
\nonumber
\theta_i|A &\stackrel{\mathit{ind}}\sim& N(\mathbf{x}_i^T\bbe, A),\quad i=1,\ldots, m.
\end{eqnarray}

In the above $\mathbf{x}_1,\ldots, \mathbf{x}_m$
are $p$-dimensional design vectors and
$\bbe$ $(p\times1)$ is the unknown regression coefficient. Writing
$\theta_i=\mathbf{x}_i^T\bbe+u_i$
$(i=1,\ldots, m),$ it is easy to reexpress (\ref{e2p1}) as a random
effects model with
%
%
\begin{equation}
y_i=\mathbf{x}_i^T\bbe+u_i+e_i,\quad i=1,\ldots, m,
\label{e2p2}
\end{equation}
where the $u_i$ and the $e_i$ are mutually\vspace*{-2pt} independent with
$u_i\stackrel{\mathit{i.i.d.}}\sim N(0, A)$
and the $e_i\stackrel{\mathit{ind}}\sim N(0, V)$. Further, writing
$\bX=(\mathbf{x}
_1,\ldots,\mathbf{x}_m)^T$,
$\mathbf{y}=(y_1,\ldots,y_m)^T$, $\bu=(u_1,\ldots,
u_m)^T$ and $\be=(e_1,\ldots,
e_m)^T$, one can rewri\-te~(\ref{e2p2}) in matrix notation as
%
%
\begin{equation}
\mathbf{y}=\bX\bbe+\bu+\be.
\label{e2p3}
\end{equation}
We assume $\operatorname{rank}(\bX)=p (<m)$. Noting that margi\-nally,
$\mathbf{y}\sim
N(\bX\bbe,(V+A)\bI_m)$, where $\bI_m$ is the identity matrix of
order $m$,
it is clear that we encounter an identifiability problem when both $V$
and $A$
are unknown. The problem does not occur in a unit-level model when one can
find a separate estimate of $V$ by utilizing the unit-level data. However,
this option is unavailable in an area-level model, where it is customary
to assume a known $V$. In practice,~$V$ is a sort of smoothed estimate,
for example, using the generalized variance function approach; see, for example,
Wolter (\citeyear{Wol85}) or Otto and Bell (\citeyear{Ot95}).

First assume $A(>0)$ is known. We begin with the HB model with the
prior $\pi(\bbe)=1$. Then we have
the following theorem.

\begin{theorem}\label{th1}
Under the given model, the posterior distribution of $\bth$ is
$N((1-B)\mathbf{y}+B\bP_{\bX}\mathbf{y},$ $
V((1-B)\bI_m+B\bP_{\bX}))$, where
$B=V/(V+A)$ and
$\bP_{\bX}=\bX(\bX^T\bX)^{-1}\bX^T$.
\end{theorem}

\begin{pf}The result follows by noting that $\bth|\bbe,\break\mathbf{y}\sim
{N}((1-B)\mathbf{y}+B\bX\bbe, V(1-B)\bI_m)$ and $\bbe
|\mathbf{y}\sim {N}((\bX^T\bX)
^{-1}\bX^T\mathbf{y},(V+A)(\bX^T\bX)^{-1})$, and then
using the formulas for
iterated expectation and variance along with normality of the conditionals.\vadjust{\goodbreak}
\end{pf}

\begin{remark}\label{rem1}It follows from the above theorem that the posterior
mean given by
%
%
\begin{equation}
\hat{\bth}{}^B= E(\bth|\mathbf{y})=(1-B)\mathbf{y}+B\bP_{\bX}\mathbf{y}
\label{e2p4}
\end{equation}
is a weighted average of the direct estimator $\mathbf{y}$
and the regression
estimator $\bP_{\bX}\mathbf{y}=
\bX\hat{\bbe}$, where $\hat{\bbe}=\break (\bX^T\cdot \bX)^{-1}\bX^T\mathbf{y}.$ It is easy
to check that the weights are
inversely proportional to the sample variance and the prior variance. Thus
$\hat{\bth}{}^B$ shrinks the direct estimator $\mathbf{y}$
of $\bth$ to the
regression estimator $\bP_{\bX}\mathbf{y}$ of $\theta$,
where the amount of
shrinking depends on the ratio $V/A$. In the limiting cases
when $B\rightarrow0$ (i.e., when $V\ll A$) or $B\rightarrow1$
(i.e., when $V\gg A$),~$\hat{\bth}{}^B$ tends respectively
to the direct estimator $\mathbf{y}$ and the regression
estimator $\bP_{\bX}\mathbf{y}$,
quite in keeping with\break one's intuition. Later, in Theorem \ref{th2},
we will motivate
the estimator in (\ref{e2p4}) as the BLUP of $\bth$ without any distributional
assumption. We also point out that this is also the best unbiased predictor
under normality.\vspace*{-1pt}
\end{remark}

\begin{remark}\label{rem2} $\!\!\!$It is also important to note that~if~the
parameter $\bbe$ were also known, the posterior variance of $\bth$
would be
$V(1-B)\bI_m$. Thus the term~$VB\bP_{\bX}$ in the posterior variance in
Theorem \ref{th1} can be interpreted as the additional posterior uncertainty
due to unknown $\bbe$, but known $A$. We will
examine later in this section the effect of an unknown $A$ as well on
the posterior variance.\vspace*{-1pt}
\end{remark}

Next we show that the estimator of $\bth$ given in~(\ref{e2p4}) can be
motivated without any distributional
assumption but using only the first two moments. The following theorem
proves that this estimator is a BLUP, that is, it
has the smallest mean squared error (MSE)
within the class of all linear unbiased estimators (predictors) of
$\bth$.
Also, the MSE equals the posterior variance given in Theorem \ref{th1}.\vspace*{-1pt}

\begin{theorem}\label{th2}
$\!\!\!$The estimator $\hat{\bth}{}^B$ of $\bth$ given in~\textup{(\ref{e2p4})} is the
BLUP of
$\bth$. Also, $E[(\hat{\bth}{}^B-\bth)(\hat{\bth}{}^B-\bth)^T]=
V\{(1-B)\bI_m+B\bP_{\bX}\}$.\vspace*{-1pt}

\end{theorem}

\begin{pf} Since $E(\bY)=\bX\bbe$, any linear unbiased
predictor $\bC\bY+\mathbf{b}$
of $\bth=\bX\bbe+\bu$ must satisfy $\bC\bX\bbe+ \mathbf{b}=\bX\bbe$ for all
$\bbe$. That is, $\mathbf{b}=\bdzero,$
and $\bC\bX=\bX$, or equivalently, $\bC\bP_{\bX}=\bP_{\bX}$.
For such a
predictor $\bC\bY$, since $\bC\bY-\bth=(\bC-\bI_m)\bu+\bC\be$,
%
%
\begin{eqnarray}\label{blup1}
&&E[(\bC\bY-\bth)(\bC\bY-\bth)^T] \nonumber\\[-0.5pt]
&&\quad= A(\bC-\bI_m)(\bC-\bI
_m)^T+V\bC\bC
^T\nonumber\\[-0.5pt]
&&\quad= (V+A)\bC\bC^T-A(\bC+\bC^T)+A\bI_m
\nonumber
\\[-8pt]
\\[-8pt]
\nonumber
&&\quad= V(1-B)\bI_m \\[-0.5pt]
&&\qquad{}+ (V+A)\{\bC-(1-B)\bI_m\}\nonumber\\[-0.5pt]
&&\qquad{}\cdot\{\bC-(1-B)\bI_m\}^T.\nonumber
\end{eqnarray}
Now subject to the condition $\bC\bP_{\bX}=\bP_{\bX}$, it can be
shown that
%
%
\begin{eqnarray}\label{blup2}
&&\{\bC-(1-B)\bI_m\}\{\bC-(1-B)\bI_m\}^T \nonumber\\[-0.5pt]
&&\quad= \{\bC-(1-B)\bI_m-B\bP_{\bX}+B\bP_{\bX}\}\nonumber\\[-0.5pt]
&&\qquad{}\cdot\{\bC-(1-B)\bI
_m-B\bP_{\bX
}+B\bP_{\bX}\}^T\\[-0.5pt]
&&\quad=\{\bC-(1-B)\bI_m-B\bP_{\bX}\}\nonumber\\[-0.5pt]
&&\qquad{}\cdot\{\bC-(1-B)\bI_m-B\bP_{\bX}\}
^T+B^2\bP
_{\bX}.\nonumber
\end{eqnarray}
Note that $\bC=(1-B)\bI_m+B\bP_{\bX}$ satisfies the condition $\bC
\bP
_{\bX}=\bP_{\bX}$ and this choice
minimizes $E[(\bC\bY-\bth)(\bC\bY-\bth)^T]$. Thus the BLUP of
$\bth$ is
given by $\hat{\bth}{}^B$. Also,
from (\ref{blup1}) and (\ref{blup2}), it follows that
the mean squared and product matrix of prediction error of the BLUP is
$V\{(1-B)\bI_m+B\bP_{\bX}\}$.
\end{pf}

\begin{remark}\label{rem3} Under normality of $\bu$ and $\be$, the BLUP
$\hat
{\bth
}{}^B$ of $\bth$ is also the
best unbiased predictor of $\bth$; that is, among all unbiased predictors
of $\bth$, $\hat{\bth}{}^B$ has the least mean squared error.
\end{remark}

Theorems \ref{th1} and \ref{th2} establish the equivalence of the BLUP and the HB predictor
and also of the corresponding uncertainty measures
for the balanced one-way random effects model when the parameter~$A$ is known.
Indeed, the result is also true for the general mixed effects model
(see, e.g.,
Datta, \citeyear{Dat92}). However, this algebraic equality does not
quite hold for unknown
$A$, or equivalently unknown $B$.

To see this, we will consider separately, the EBLUP (or EB) and HB estimators,
and point out where the differences occur. For the given random effects model,
$\mathbf{y}\sim N(\bX\bbe, (V+A)\bI_m)$, which in
the Baye\-sian terminology, is the marginal distribution of $\mathbf{y}$ after
integrating out $\bth$. Based on this
marginal pdf, $(\hat{\bbe}, S=\Vert \mathbf{y}-\bX\hat{\bbe
}\Vert ^2)$ is minimal
sufficient for $(\bbe, A)$. Noting that
$S\sim(V+A)\chi^2_{m-p},$ the UMVUE of $B=V/(V+A)$ is given by ${\hat
B}^{\mathit{EB}}=V(m-p-2)/S$ for $m > p+2$.
The corresponding EB or EBLUP estimator of $\bth$ is then given by
%
%
\begin{eqnarray}\label{e2p8}
\hat{\bth}{}^{\mathit{EB}} &\equiv&\hat{\bth}{}^{\mathit{EBLUP} }
\nonumber
\\[-0.5pt]
&=&\biggl[1-\frac
{V(m-p-2)}S\biggr]\mathbf{y}\\[-0.5pt]
&&{}+\frac{V(m-p-2)}S \bX\hat{\bbe},\nonumber
\end{eqnarray}
the James--Stein estimator (James and Stein,
\citeyear{JamSte61}).\vadjust{\goodbreak}

One criticism of the above EB or EBLUP estima\-tor is that the estima\-tor
${\hat B}^{\mathit{EB}}$ of $B$ can assume~values bigger than 1 with positive
probability. The \mbox{resulting} EB or EBLUP estimator then pulls the direct
estimator $\mathbf{y}$ toward the opposite direction of the
regression estimator
$\bP_{\bX}\mathbf{y}$. Replacing\vspace*{1pt} ${\hat B}^{\mathit{EB}}$ by
$({\hat B}^{\mathit{EB}})^+$, where $({\hat B}^{\mathit{EB}})^+=\operatorname{min}({\hat B}^{\mathit{EB}},1)$,
the positive part Stein estimator rectifies the problem. However, it
was\break
shown by Datta et al. (\citeyear{Datetal02}) that $P({\hat B}^{\mathit{EB}}>1)$
goes to zero at
an exponential rate for large $m$. So, the estimator ${\hat B}^{\mathit{EB}}$ is
usually quite adequate even for moderate $m$.

In contrast, with the alternative fully Bayesian~ap\-proach (Morris,
\citeyear{Mor83N1}), if
one assigns the prior $\pi(\bbe,\break A)=1$ so that $\pi(\bbe,B)=B^{-2}$,
one gets
$\pi(\bth| B,\mathbf{y})$ the same as given in Theorem \ref{th1}
for a known $B$, but
needs in
addition
%
\begin{eqnarray*}
\pi(B|\mathbf{y})&\propto& B^{{(m-p)}/2}\exp\biggl(-\frac1{2V}
BS\biggr)B^{-2}I[0<B<1]\\
&=& B^{{(m-p-4)}/2}\exp\biggl(-\frac1{2V} B S\biggr)I[0<B<1].
\end{eqnarray*}
Here, for the sake of simplicity and to present Morris's results, we
have considered only a uniform prior
for $A$. It is certainly possible to consider other priors, including
inverse gamma priors with appropriate shape and scale parameters of the
inverse gamma distribution, so long as the resulting posterior is
proper. A prior of the form $\pi(\bbe,A)=A^{-k}$ will yield a proper posterior
provided $k\,{<}\,1$ and \mbox{$m\,{>}\,p\,{-}\,2k\,{+}\,2$}. Thus, while the uniform prior $\pi(\bbe
,A)=1$ yields a~proper posterior when $m>p+2$, the priors $A^{-1}$ or
$A^{-2}$ will always yield improper posteriors. For the uniform prior,
the posterior mean of $\bth$ is now obtained by replacing $B$ in
Theorem~\ref{th1} with $E(B|\mathbf{y})$, while $V(\bth|\mathbf{y})=V[(1-E(B|\mathbf{y}))\bI
_m+E(B|\mathbf{y})P_{\bX}]+ V(B|\mathbf{y})(\mathbf{y}-P_{\bX}\mathbf{y})(\mathbf{y}-P_{\bX}\mathbf{y})^T$. Thus,
other than the replacement of $B$ by $E(B|\mathbf{y})$ in
the variance formula
given in Theorem~\ref{th1}, the
additional uncertainty due to estimation of $B$ is also incorporated in
this variance formula.

Integrating by parts,
one can show that for large~$m$, $E(B|\mathbf{y})$ can be
approximated by
$(m-p-2)V/S$ (cf. Theorem 1 of Datta and Ghosh, \citeyear{DatGho91N1}).
Similarly,
$V(B|\mathbf{y})$ can be approximated by $2(m-p-2)V/S^2$.
With these
approximations, $E(\bth|\mathbf{y})$ is approximated by
$\hat{\bth}{}^{\mathit{EB}}$,
while $V(\bth|\mathbf{y})$ can be approximated as
$V(1-\frac{(m-p-2)V}S)\bI_m
+\frac{(m-p-2)V^2}S \bP_{\bX}+\frac{2(m-p-2)V^2}{S^2}(\mathbf{y}-\break\bX\hat{\bbe
})(\mathbf{y}-\bX\hat{\bbe})^T$.\vspace*{1pt} These results agree with
Morris' (\citeyear{Mor83N2})
intuitive approximations for
$E(\bth|\mathbf{y})$ and\break $V(\bth|\mathbf{y})$
for the special case of intercept model.
In addition, if instead of the posterior mean, one estimates $B$ by its
posterior mode, one gets the estimator $\hat{B}^{MO}=\operatorname
{min}((m-p-4)V/S,1)$, which
leads to an estimator of $\bth$ quite akin to the positive part James--Stein
estimator, the only difference being that $m-p-2$ is now replaced by
$m-p-4$.

It is instructive to find the Bayes risk of $\hat{\bth}{}^{\mathit{EB}}$ under squared
error loss
$L(\bth,\mathbf{a})=\Vert \bth-\mathbf{a}\Vert ^2$. The
following theorem is proved.

\begin{theorem}\label{th3}
Let $m > p+2$. Then writing $h_{ii}=\mathbf{x}_i^T(\bX^T\bX
)^{-1}\mathbf{x}_i$ for
all $i$:
\begin{longlist}
\item[(a)]
\[E[(\theta_i-\hat{\theta}_i^{\mathit{EB}})]^2=V(1-B)+VBh_{ii}+
\frac{2VB(1-h_{ii})}{m-p};
\]
\item[(b)]
\[
E\Vert \bth-\hat{\bth}^{\mathit{EB}}\Vert ^2=V[m-(m-p-2)B].
\]
\end{longlist}
\end{theorem}

\begin{pf} Let $\hat{B}=V(m-p-2)/S$.
If $\tilde{\theta}_i=E[\theta_i| \bbe,\break A,\mathbf{y}]$,
then $\tilde{\theta
}_i=y_i-B(y_i-\mathbf{x}_i^T\bbe)$ and $V[\theta_i|\bbe
,A, \mathbf{y}]
=V(1-B).$ Using iterated expectation it follows that
%
%
\begin{equation}
\hspace*{28pt} E[(\theta_i-\hat{\theta}_i^{\mathit{EB}})]^2=V(1-B)+E[(\tilde{\theta
}_i-\hat
{\theta}_i^{\mathit{EB}})]^2 .
\label{eg1}
\end{equation}
Using the expressions of $\tilde{\theta}_i$, $\hat{\theta}_i^{\mathit{EB}}$, and
independence of $\hat{\bbe}$ and $\mathbf{y}-\bX\hat
{\bbe}$,
it follows that
%
%
\begin{eqnarray}\label{eg2}
&& E[(\tilde{\theta}_i-\hat{\theta}_i^{\mathit{EB}})]^2\nonumber\\
&&\quad= E[\{B\mathbf{x}_i^T(\hat{\bbe
}-\bbe)\}^2]
\nonumber
\\[-8pt]
\\[-8pt]
\nonumber
&&\qquad{}+ E[(\hat{B}-B)^2(y_i-\mathbf{x}_i^T\hat
{\bbe})^2]\\
&&\quad= VBh_{ii} +
E[(\hat{B}-B)^2(y_i-\mathbf{x}_i^T\hat{\bbe})^2],\nonumber
\end{eqnarray}
where $h_{ii}=\mathbf{x}_i^T(\bX^T\bX)^{-1}\mathbf{x}_i$.
By Basu's theorem, $S$ and $(y_i-\mathbf{x}_i^T\hat{\bbe
})^2/S$ are
independently distributed (see Ghosh, \citeyear{Gho92N1}). Then
%
%
\begin{eqnarray}\label{eg3}
&& E[(\hat{B}-B)^2(y_i-\mathbf{x}_i^T\hat{\bbe})^2]
\nonumber
\\[-8pt]
\\[-8pt]
\nonumber
&&\quad=E[S(\hat
{B}-B)^2]E[(y_i-\mathbf{x}
_i^T\hat{\bbe})^2/S].
\end{eqnarray}
By a simple calculation $E[S(\hat{B}-B)^2]=2VB$. Also, by the
independence of
$S$ and $(y_i-\mathbf{x}_i^T\hat{\bbe})^2/S$,
%
%
\begin{eqnarray}\label{eg4}
E[(y_i-\mathbf{x}_i^T\hat{\bbe})^2/S]&= &\frac{E(y_i-\mathbf{x}_i^T\hat{\bbe})^2}{E(S)}
\nonumber\\
&=&\frac{(\sigma^2/B)(1-h_{ii})}{(\sigma^2/B)(m-p)}
\\
&=&\frac{1-h_{ii}}{m-p}.\nonumber
\end{eqnarray}
Combining (\ref{eg1})--(\ref{eg4}), one gets (a). Summing\break both sides
of (a)
over $i$, and noting
$\sum_{i=1}^m h_{ii}=\break\operatorname{tr}[(\bX^T\bX)^{-1}\cdot (\bX^T\bX)]=p$, one
gets (b).
\end{pf}

\begin{remark}\label{rem4} $\!\!\!$It is interesting to observe that a~comparison of
Theorem \ref{th3} with Theorem~\ref{th1} (or Theorem~\ref{th2})
reveals that the excess Bayes risk due to estimation of the unknown
variance component $A$ is simply $2VB$. It is easy
to see from Theorem \ref{th1} or \ref{th2} that the Bayes risk with known $A$ is
$V[m(1-B)+pB]$.
\end{remark}

\begin{remark}\label{rem5} $\!\!\!$Another interesting observation from Theorem \ref{th3} is that
an unbiased estimator of the MSE\vspace*{1pt} is $V[m-\frac{V(m-p-2)^2}S]$
which is simply Stein's\vspace*{1pt} unbiased estimator. While this is in agreement
with equation (1.18) of Morris (\citeyear{Mor83N2}), our expression
for the component MSE given by part (a) in Theorem \ref{th3} agrees with
equation (1.16) of Morris
(\citeyear{Mor83N2}) only in the special case of an intercept model,
that is, when
$\theta_i=\mu+u_i$ ($i=1,\ldots, m$). We believe
that this is due to an oversight in Morris (\citeyear{Mor83N2}) in the
derivation of the
component risk for the general regression model.
\end{remark}

We will now see how the above results can be generalized with unequal numbers
of observations in the different small areas.

\section{Shrinkage Estimators for Unbalanced Data}\label{sec3}
The equal sampling variance scenario considered in the previous section
hardly arises for small area problems, where sampling
variances for small areas are almost always unequal. A widely used
area-level model first introduced by Fay and Herriot (\citeyear
{FayHer79}) is
given by
%
%
\begin{equation}
y_i|\theta_i \stackrel{\mathit{ind}}\sim N(\theta_i, V_i), \quad\theta
_i\stackrel{\mathit{ind}}\sim N(\mathbf{x}_i^T\bbe, A).
\label{e3p1}
\end{equation}
Clearly the above model can be viewed also as a random effects model as
shown in the previous section.

Fay and Herriot used the above model for estimating the per capita
income (PCI) for small places in the United States with population
less than 1000. In their case, $y_i$ is the logarithm of per capita
income for the $i$th small area. The auxiliary variables
considered were logarithms of the PCI for the associated counties, tax
return data, data on housing from the previous decennial census.
The Fay--Herriot method was adopted by the U. S. Bureau of the Census
to provide updated PCI estimates for small areas.

Fay and Herriot adopted an EB approach in their analysis. Write\vadjust{\goodbreak} $\bG
=\operatorname{Diag}(V_1,\ldots, V_m),$
$ \bD=\bG+A\bI_m$, $\bB=\bG{\bD}^{-1}={\bD}^{-1}\bG= \operatorname
{Diag}(B_1,\ldots,B_m),$ where\break $B_i=V_i/(V_i+A),
i=1,\ldots, m$. First, assuming $\bbe$ and $A$ to be both known, the Bayes
estimator of $\bth$ is $\hat{\bth}{}^B = (\bI_m-\bB)\mathbf{y}+\bB\bX\bbe$.
In order to estimate $\bbe$ and~$A$ as needed in an EB approach, first
observe that for $A$ known, the
generalized least squares estimator of $\bbe$ is
%
%
\begin{eqnarray}\label{e3p2}
\tilde{\bbe}(A)&=&(\bX^T\bD^{-1}\bX)^{-1}\bX^T\bD^{-1}\mathbf{y}
\nonumber
\\[-8pt]
\\[-8pt]
\nonumber
&=&[\bX^T(\bI_m-\bB
)\bX]^{-1}\bX^T(\bI_m-\bB)\mathbf{y},
\end{eqnarray}
where we assume, as before, rank$(\bX)=p (< m).$
We may note here that the corresponding\vspace*{1pt} BLUP estimator of $\bth$ is
$(\bI_m-\bB)\mathbf{y}+\bB\bX\tilde{\bbe}(A)$. In
order to estimate~$A$ as well, Fay and Herriot (\citeyear{FayHer79}) and Datta, Rao and Smith
(\citeyear{DatRaoSmi05})\vspace*{1pt} used the
moment identity given by
$E[\sum_{i=1}^m\{y_i-\mathbf{x}_i^T\tilde{\bbe}(A)\}
^2/(V_i+A)] = m-p.
$ Dropping the expectation from the left-hand side we get
%
%
\begin{equation}
\sum_{i=1}^m\{y_i-\mathbf{x}_i^T\tilde{\bbe}(A)\}
^2/(V_i+A) = m-p.
\label{e3p3}
\end{equation}
Since the expression in the left-hand side of (\ref{e3p3}) is a~nonincreasing function of $A$, if this expression evaluated at $A=0$ is
less than $m-p$, there will be no solution to the above equation. In
this case, the estimate is taken to be zero. In the other case, taking
an initial guess at $A$ and solving~(\ref{e3p2}) and~(\ref{e3p3})
iteratively, one finds the estimators~$\hat{A}$ and~$\hat{\bbe}=\tilde{\bbe}(\hat{A}).$ The resulting EB or EBLUP estimator
of $\bth$ is given by
%
%
\begin{equation}
\hat{\bth}{}^{\mathit{EB}}=(\bI_m-\hat{\bB})\mathbf{y}+\hat{\bB
}\bX\hat{\bbe},
\label{e3p4}
\end{equation}
where $\hat{\bB}=\operatorname{Diag}({V_1}/({V_1+\hat{A}}),\ldots,
{V_m}/({V_m+\hat{A}})).$

Morris (\citeyear{Mor83N2}) provided a general discussion of the EB
approach in this
case with the same prescription for estimation of
$\bbe$ and $A$. An alternative HB formulation analogous to the one in
Section \ref{sec2} is given by Ghosh (\citeyear{Gho92N1}) who also
explored an
interrelationship between the EB and the HB procedures. The HB model is
given by
%
%
\begin{eqnarray}\label{e3p5}
&&y_i|\theta_i,\bbe, A \stackrel{\mathit{ind}}\sim N(\theta_i,V_i),\nonumber\\
&&\theta
_i|\bbe,A \stackrel{\mathit{ind}}\sim N(\mathbf{x}_i^T\bbe, A),
\\
&&\quad i=1,\ldots, m,\quad
\pi(\bbe,A)=1 .\nonumber
\end{eqnarray}
Then the joint posterior density is
%
%
\begin{eqnarray}\label{e3p6}
&&\pi(\bth, \bbe, A|\mathbf{y}) \nonumber\\
&&\quad\propto A^{- m/2}\exp
\bigl[-\tfrac12 \{(\mathbf{y}-\bth
)^T\bG^{-1}(\mathbf{y}-\bth)
\\
&&\hspace*{96pt}{} + A^{-1}\Vert \bth-\bX\bbe
\Vert ^2\}\bigr] .\nonumber
\end{eqnarray}
Then one gets $\bth|\bbe,a,\mathbf{y}\sim{N}[(\bI
_m-\bB)\mathbf{y}+\bB\bX\bbe,\break
\bG(\bI_m-\bB)]$, $\bbe|A,\mathbf{y}\sim
{N}[\tilde{\bbe}(A),
A\{\bX^T(\bI_m-\bB)\bX\}^{-1}]$, where $\tilde{\bbe}(A)=[\bX
^T(\bI_m-\bB
)\bX]^{-1}\bX^T(\bI_m-\bB)\mathbf{y}$.
The marginal posterior of $A$ is
%
%
\begin{eqnarray}\label{e3p11}
\pi(A|\mathbf{y})&\propto& A^{-{(m-p)}/2}\prod
_{i=1}^m(1-B_i)^{1/2}
\nonumber
\\[-8pt]
\\[-8pt]
\nonumber
&&{}\cdot\Biggl|
\sum_{i=1}^m(1-B_i)\mathbf{x}_i\mathbf{x}_i^T\Biggr |^{-1/2} \exp\biggl[-\frac12 Q(\mathbf{y})\biggr],
\end{eqnarray}
where $Q(\mathbf{y})\!=\!A^{-1}[\sum_{i=1}^m (1\!-\!B_i)y_i^2\!-\!
\{\sum_{i=1}^m(1\!-\!\break B_i)y_i\mathbf{x}_i\}^T \{\sum
_{i=1}^m(1\!-\!B_i)\mathbf{x}_i\mathbf{x}_i^T\}
^{-1}\{\sum_{i=1}^m(1\!-\!B_i)y_i\mathbf{x}_i\}]$.
It follows now that
%
%
\begin{eqnarray}
E(\bth|\mathbf{y})&=&[\bI_m- E(\bB|\mathbf{y})]\mathbf{y}+ E[\bB H_{\bX}\mathbf{y}|\mathbf{y}], \hspace*{-25pt}\label{e3p12}
\\
\label{e3p13} V(\bth|\mathbf{y})&=&E[\{\bI_m-\bB\}\bG|\mathbf{y}]\nonumber\\
&&{}+E[\{\bB(\bI_m-H_{\bX})\}\bG|\mathbf{y}]\hspace*{-25pt}
\\
&&{}+
V[\bB\{\mathbf{y}-\bX\tilde{\bbe}(A)\}|\mathbf{y}],\nonumber\hspace*{-25pt}
\end{eqnarray}
where $H_{\bX}=\bX[\bX^T(\bI_m-\bB)\bX]^{-1}\bX^T(\bI_m-\bB)$.
Numerical
integration involving one-dimensional integrals needs to be carried out for
evaluating both $E(\bth|\mathbf{y})$ and $V(\bth|\mathbf{y})$. In the special case, when
$V_1=\cdots=V_m$, these expressions simplify to the ones obtained in the
previous section. This is because in this special case, $\bI_m-\bB
=(1-B)\bI_m$.
We may also reemphasize that the first component in the right-hand side of
(\ref{e3p13}) is the posterior variance when both $\bbe$ and $A$ are known.
The second term provides additional uncertainty due to unknown $\bbe$
but known $A$. The third term accounts for additional uncertainty due to
unknown $A$ as well.

In the Bayesian framework, posterior variances are the natural uncertainty
measures. In the frequentist approach, a naive method is to substitute
$A$ by some suitable estimator in the mean squared prediction error
formula for the BLUP [cf. (\ref{blup1}) and (\ref{blup2})
for the balanced case]. In the unbalanced case, the mean squared and product
matrix of prediction error for the BLUP is given by the sum of the
first two
terms in the right-hand side of (\ref{e3p13}) without the conditional
expectation operator.
As it appears, this will miss the third component as it will not
account for
uncertainty due to estimation of~$A$. This results in an underestimation
of the true MSE of the EBLUP.

To account for the error in estimating $A$, following an earlier work
of Kackar and Harville (\citeyear{KacHar84}), Prasad and Rao
(\citeyear
{PraRao90})
considered the MSE of the EBLUP. Unlike in the balanced case of Section
\ref{sec2}, there is no closed-form expression of this MSE.
They obtained an asymptotic expression of the MSE which is accurate to
the order $o(m^{-1})$. This
approximation is based on an orthogonal decomposition of the MSE
$E(\hat{\theta}_i^{\mathit{EB}}-\theta_i)^2$. Specifically, they used the
decomposition
\begin{eqnarray*}
\hat{\theta}_i^{\mathit{EB}}-\theta_i &=& \{(1-B_i)y_i+B_i\mathbf{x}_i^T\bbe- \theta_i
\}\\
&&{} + \{\hat{\theta}_i^{B}-(1-B_i)y_i-B_i\mathbf{x}_i^T\bbe
\}\\
&&{}+ \{\hat{\theta}_i^{\mathit{EB}}-\hat{\theta}_i^{B}\};
\end{eqnarray*}
the first component is always orthogonal to the second and the third
components. The
orthogonality of the last two components holds only for certain
specific estimators of $A$. It is necessary that these estimators
are translation invariant under the transformation $g_c(\mathbf{y})$ which
maps $\mathbf{y}$
to $\mathbf{y}+\bX\mathbf{c}$ and are even
functions of $\mathbf{y}$.
In particular, the ANOVA estimator in Prasad and Rao (\citeyear
{PraRao90}), the ML and
the REML estimators
considered in Datta and Lahiri (\citeyear{DatLah00}), and the method
of moment
estimator due to Fay and Herriot (\citeyear{FayHer79}), Morris
(\citeyear{Mor83N2}) and
Datta, Rao and Smith (\citeyear{DatRaoSmi05}) all satisfy these conditions. For
these estimators
of $A$,
it follows that
%
%
\begin{eqnarray}\label{e3p14}
\qquad\quad&&E(\hat{\theta}_i^{\mathit{EB}}-\theta_i)^2
\nonumber
\\[-8pt]
\\[-8pt]
\nonumber
&&\quad= g_{1i}(A)+g_{2i}(A)+g_{3i}(A)+o(m^{-1}),
\end{eqnarray}
where $g_{1i}(A)=V_i(1-B_i)$, $g_{2i}(A)=B_i^2A\mathbf{x}_i^T\cdot\break \{\sum_{j=1}^m
(1-B_j)\mathbf{x}_j\mathbf{x}_j^T\}^{-1}\mathbf{x}_i$ and $g_{3i}(A)=V_i^2(A+\break  V_i)^{-3}
\operatorname{Var}(\hat{A})$. The derivation of the third term is~ba\-sed on a
second-order Taylor expansion [i.e., retaining up to the $O(m^{-1})$
term] of
$E[B_i(y_i-\mathbf{x}_i^T\tilde{\bbe}(A))-\hat
{B}_i(y_i-\mathbf{x}_i^T\tilde{\bbe
}(\hat{A}))]^2$.
This derivation requires also orthogonality of $\bbe$ and $A$ in the
Fisherian sense, that is, block diagonality of the relevant components
of the
Fisher information matrix.
An intuitive estimator, say, $\mathit{mse}^I(\hat{A})$, of the MSE in (\ref{e3p14})
is given by
%
%
\begin{equation}
\mathit{mse}^I(\hat{A})=g_{1i}(\hat{A})+g_{2i}(\hat{A})+g_{3i}(\hat{A}).
\label{mseI}
\end{equation}
In view of the fact that $E[g_{1i}(\hat
{A})]=g_{1i}(A)-g_{3i}(A)+o(m^{-1})$, and
$g_{3i}(A)$ is $O(m^{-1})$, the above estimator is not second-order unbiased.
Based on the ANOVA estimator of $A$, say, $\hat{A}_{\mathit{PR}}$, which is
second-order unbiased for $A$, Prasad and Rao (\citeyear{PraRao90})
showed that the estimator
%
%
\begin{equation}
\mathit{mse}^S(\hat{A}_{\mathit{PR}})=\hat{g}_{1\mathit{PR}i}+\hat{g}_{2\mathit{PR}i}+\hat{g}_{3\mathit{PR}i}
\label{mse2}
\end{equation}
is second-order unbiased in the sense that
\[
E[\mathit{mse}^S (\hat{A}_{\mathit{PR}})]=
\mathit{MSE}(\hat{\theta}_i^{\mathit{EB}})+o(m^{-1}),
\]
where
\begin{eqnarray*}
\hat{g}_{1\mathit{PR}i}&=&g_{1i}(\hat{A}_{\mathit{PR}})+g_{3i}(\hat{A}_{\mathit{PR}}),\quad \hat
{g}_{2\mathit{PR}i}=g_{2i}(\hat{A}_{\mathit{PR}}), \\
\hat{g}_{3\mathit{PR}i}&=&g_{3i}(\hat{A}_{\mathit{PR}}).
\end{eqnarray*}
See Harville (\citeyear{Har90}) for similar results for mixed linear
models. In the
small area context Datta and Lahiri (\citeyear{DatLah00})
showed that the expression in (\ref{mse2}) based on the REML estimator
of $A$ is also second-order unbiased. Second-order
unbiased estimator of the MSE of the EBLUP using the ML estimator and
Fay--Herriot estimator of $A$ are given
in Datta and Lahiri (\citeyear{DatLah00}) and Datta, Rao and Smith (\citeyear
{DatRaoSmi05}), respectively. For further
discussion we may refer to Rao (\citeyear{Rao03N1}) and Datta
(\citeyear{Dat}).

The posterior variance of $\theta_i$, on the other hand [see~(\ref{e3p13})],
is given by
%
%
\begin{eqnarray}\label{e3p15}
V(\theta_i|\mathbf{y}) &=& V_i[1-E(B_i|\mathbf{y})] \nonumber\\
&&{}+ E\Biggl[B_i^2A\mathbf{x}_i^T\Biggl\{\sum
_{j=1}^m(1-B_j)\mathbf{x}_j\mathbf{x}_j^T\Biggr\}
^{-1}\mathbf{x}_i\Big|\mathbf{y}\Biggr]\nonumber\nonumber\\
&&{} + V[B_i\{y_i-\mathbf{x}_i^T\tilde{\bbe}(A)\}|\mathbf{y}]
\nonumber
\\[-8pt]
\\[-8pt]
\nonumber
&=& E[g_{1i}(A)|\mathbf{y}]+ E[g_{2i}(A)|\mathbf{y}]\\
&&{}+ V[B_i\{y_i-\mathbf{x}_i^T\tilde{\bbe
}(A)\}|\mathbf{y}]\nonumber\\
&=& g_{1\mathit{HB}i}+g_{2\mathit{HB}i}+g_{3\mathit{HB}i} \quad\mbox{(say)}.
\nonumber
\end{eqnarray}

Morris (\citeyear{Mor83N2}) provided an approximation to the HB
estimator $E(\theta
_i|\mathbf{y})$ and the associated posterior\vspace*{-1pt}~va\-riance.
Denoting Morris' point
estimator of $\theta_i$ by~$\hat{\theta_i}^M$,
%
%
\begin{equation}
\hat{\theta_i}{}^M = (1-{\hat B}_i^M)y_i+{\hat B}_i^M(\mathbf{x}_i^T \hat{\bbe}),
\label{morris-mean}
\end{equation}
where ${\hat B}_i^M=((m-p-2)/(m-p))(V_i/(V_i+{\hat A }))$, and~$\hat
{\bbe}$ and $\hat A$ are obtained by solving (\ref{e3p2}) and (\ref
{e3p3}) iteratively.
It can be checked that (\ref{e3p2}) and (\ref{e3p3}) are equivalent
to Morris'
(\citeyear{Mor83N2}) equations~(5.2) and~(5.4). Morris (\citeyear
{Mor83N2}) approximated the posterior
variance by $s_{iM}^2$, given by $s_{iM}^2=e_{iM}+v_{iM}$,
where $e_{iM}=g_{1iM}+g_{2iM}$, $v_{iM}=g_{3iM}$ with
%
%
\begin{eqnarray}\label{morris-var}
\qquad g_{1iM}&=& V_i[1-{\hat B}_i^M],\quad g_{2iM}=V_i {\hat B}_i^M{\hat
t}_i,
\nonumber
\\[-8pt]
\\[-8pt]
\nonumber
g_{3iM}&=& \frac{2({\hat B}_i^{M})^2(y_i-\mathbf{x}_i^T \hat
{\bbe})^2}{m-p-2}
\cdot\frac{{\bar V}+{\hat A}}{V_i+{\hat A}},
\end{eqnarray}
and ${\hat t}_i=\mathbf{x}_i^T [\bX^T(\bV+{\hat A}\bI
)^{-1}\bX]^{-1}\mathbf{x}_i/(V_i+{\hat A})$, $i=1,\break\ldots, m$, $\bar V = \sum_{i=1}^mV_i/m$.

From the three measures of uncertainty given\break by~(\ref{mse2}), (\ref
{e3p15}) and (\ref{morris-var}) we see a close correspondence
in the respective terms in the expansion of the MSE of the EB\vadjust{\goodbreak}
estimator, the posterior variance of~$\theta_i$ and Morris' approximation
of the posterior variance.
It is clear, though, that while the posterior variance of $\theta_i$ accounts
for all sources
of uncertainty in a straightforward way, the EB or EBLUP method needs careful
evaluation of
all terms in the MSE expression and construct a second-order unbiased
estimator of this quantity. Morris (\citeyear{Mor83N2}) provided a~clever approximation
to the posterior variance. The estimator of the MSE of the
EBLUP displays poor performance when $A$ is estimated by zero or
severely underestimated (this happens if the true variance parameter
$A$ is small).
In such case the first term $\hat{g}_{1\mathit{PR}}$ is too small compared to
the first term in the posterior variance. This results from the
integration of
$A$ with respect to its long tail posterior distribution. Use of posterior
variance has been found to be attractive in small area application. As an
example, the U.S. Bureau of the Census uses this method in producing
small area income and poverty estimates based on American Community Survey
data. The corresponding term in Morris' approximation is a~clever
approximation to the posterior expectation. Although not as small as
$\hat{g}_{1\mathit{PR}}$, this
also tends to be small. The $g_{1i}(A)$ function evaluated at the point
estimator of $A$, via posterior mode or REML, is
usually smaller than its integrated value with respect to the posterior
of $A$. The second and the third terms in these measures of
uncertainty, being of lower order of magnitude, usually show a greater
degree of agreement.
Another
attractive feature of posterior variance is that it depends on the
individual small area observation $y_i$ [through the
last term in (\ref{e3p15})]. This is not true for the second-order unbiased
estimator of the MSE given in (\ref{mse2}). However,
the estimate of conditional frequentist mean squared error of
prediction obtained by conditioning on $y_i$ depends on the individual
small area observation (see, e.g., Booth and Hobert, \citeyear
{BooHob98}, or Datta et
al., \citeyear{Datetal}). For related discussions comparing the Bayesian
and the frequentist measures of uncertainty in small area estimation we
refer to Singh, Stukel and Pfeffermann (\citeyear{SinStuPfe98}) and Datta, Rao and Smith
(\citeyear{DatRaoSmi05}).
Morris' approximation, which closely mimics the posterior variance,
also enjoys this feature.

\begin{table*}
\caption{Data for estimating 1,979 four-person family
median income for the 15 southeastern U.S. states, and different small
area estimates}\label{tab3.1}
\begin{tabular*}{\textwidth}{@{\extracolsep{\fill}}lccccccccc@{}}
\hline
\textbf{State} & $\bolds{y}$ & $\bolds{x}$ & $\bolds{V}$ &$\bolds{\hat{\theta}}^{\bolds{b}\bolds{\mathit{HB}}}$ &
$\bolds{\hat{\theta}}^{\bolds{b}\bolds{\mathit{EB}}}$
&$\bolds{\hat{\theta}}^{\bolds{bM}}$ &$\bolds{\hat{\theta}}^{\bolds{u}\bolds{\mathit{HB}}}$ &$\bolds{\hat{\theta}}^{\bolds{u}\bolds{\mathit{EB}}}$
&$\bolds{\hat
{\theta}}^{\bolds{uM}}$\\
\hline
DE & 21,860 & 23,103 & 1,900$^2$ & 21,185 & 21,787 & 21,031 & 21,088 & 21,802
&21,025\\
MD & 26,235 & 27,607 & 1,722$^2$ & 25,399 & 26,145 & 25,221 & 25,227 & 26,134 &
25,090 \\
VA & 24,160 & 25,514 & 1,418$^2$ & 23,418 & 24,080 & 23,264& 23,403 & 24,040 &
23,262\\
WV & 18,274 & 21,807 & 1,380$^2$ & 19,133 & 18,367 & 19,330& 19,027 & 18,397 &
19,160\\
NC & 20,296 & 21,408 & 1,012$^2$ & 19,634 & 20,223 & 19,472& 19,849 & 20,133 &
19,712\\
SC & 19,282 & 21,706 & 1,795$^2$ & 19,448 & 19,299 & 19,472& 19,452 & 19,296 &
19,454\\
GA & 22,687 & 22,599 & 1,196$^2$ & 21,217 & 22,524 & 20,842& 21,510 & 22,402 &
21,199\\
FL & 19,675 & 23,944 & 1,042$^2$ & 20,884 & 19,807 & 21,174& 20,480 & 19,941 &
20,700\\
AL & 17,978 & 22,233 & 1,282$^2$ & 19,273 & 18,119 &19,575& 19,047 & 18,187 &
19,264\\
KY & 18,657 & 21,359 & 1,285$^2$ & 19,008 & 18,695 & 19,087& 18,954 & 18,716 &
19,017\\
TN & 19,776 & 21,240 & 1,274$^2$ & 19,351 & 19,729 &19,239& 19,430 & 19,707 &
19,350\\
MS & 19,167 & 19,887 & 1,762$^2$ & 18,360 & 19,075 & 18,131& 18,371 & 19,097 &
18,274\\
AR & 18,917 & 20,214 & 1,507$^2$ & 18,388 & 18,858 & 18,250&18,452 & 18,859 &
18,383\\
LA & 18,965 & 22,861 & 1,444$^2$ & 19,996 & 19,078 & 20,240& 19,878 & 19,096 &
20,020\\
OK & 19,295 & 23,668 & 1,675$^2$ & 20,578 & 19,436 & 20,894 &20,535& 19,418 &
20,673\\
\hline
\end{tabular*}
\end{table*}

We consider an illustration of the Fay--Herriot mo\-del. The U.S.
Department of Health and Human Services (HHS) needs estimates of
four-person family state median income to implement an energy
assistance program to low-income families. The Bureau of the Census
(BOC) has provided such estimates for nearly thirty years. The BOC
now uses the Fay--Herriot model\vadjust{\goodbreak} to provide more sophisticated
estimates. In this model the direct estimate of the four-person family
state median income, to be denoted by $y_i$, is obtained from the
Current Population Survey (CPS). Auxiliary variables for the multiple
regression model are obtained from the per capita income information of
the Bureau of the Economic Analysis (BEA) survey and the latest census
data for the four-person
family median income. In our illustration, we will consider only a
subset of the U.S. states and use only one covariate. We consider 15
U.S. states belonging to the southeast U.S. geographical region. While
there are 17 states in this region, we excluded Texas and Washington,
DC, from our analysis as these two small areas have their sampling
variances ($V_i$'s) very much different from the remaining 15 states.
In the notation of this section, we have $m=15$, $p=2$, $\mathbf{x}_i^T=(1,
x_i)$, with $x_i$, the adjusted census median income, given by
\begin{eqnarray*}
x_{i}&=& \frac{\mbox{BEA PCI(c) for state $i$}} {\mbox{BEA PCI($b$) for
state $i$}}\\
&&{}\cdot\mbox{Census median(b) for state $i$},
\end{eqnarray*}
where $c$ stands for current year (in our application 1979) and $b$
stands for base year (1969), BEA PCI($b$) and BEA PCI($c$) are obtained
from the BEA data for these two years, and Census median($b$) is
obtained from the 1969 census.


We present the relevant data in Table \ref{tab3.1} below. Also included in the
table are the EB estimates ($\hat{\theta}^{b\mathit{EB}}$ in the balanced
case,
and $\hat{\theta}^{u\mathit{EB}}$ in the unbalanced case), the HB estimates
($\hat{\theta}^{b\mathit{HB}}$ in the balanced case, and $\hat{\theta
}^{u\mathit{HB}}$ in
the unbalanced case) and Morris' approximation to the HB estimates
($\hat{\theta}^{bM}$ in the balanced case, and
$\hat{\theta}^{uM}$ in the unbalanced case). As noted before, the
sampling variances are different for the states and the resulting
Fay--Herriot model is an unbalanced model. To compare the frequentist
and the Bayesian approaches for both the balanced and the unbalanced
setup, we have illustrated the balanced Fay--Herriot model given by
(\ref{e2p1}) by replacing each $V_i$ by their average 2,162,469. From
the last six columns of Table \ref{tab3.1}, we note that the point estimates of
the small area means do not differ substantially either over EBLUP, HB
or Morris' estimates, or if the setup is a balanced
or an unbalanced Fay--Herriot model. It is usually our experience that
the model-based small area point estimates are substantially robust
over varying sampling variances or over the method of estimation, Bayes
or frequentist.\looseness=1

%
\begin{table*}
\caption{Decomposition of various measures of uncertainty
of the model-based small area estimates}\label{tab3.2}
\vspace*{2pt}
\begin{tabular*}{\textwidth}{@{\extracolsep{\fill}}lcrrrrrrrrr@{}}
\hline
\textbf{State} &\textbf{Setup} &\multicolumn{1}{c}{$\bolds{g}_{\mathbf{1}\bolds{\mathit{HB}}}$} &
\multicolumn{1}{c}{$\bolds{g}_{\mathbf{2}\bolds{\mathit{HB}}}$}
&\multicolumn{1}{c}{$\bolds{g}_{\mathbf{3}\bolds{\mathit{HB}}}$}
&\multicolumn{1}{c}{$\bolds{g}_{\bolds{1M}}$}& \multicolumn{1}{c}{$\bolds{g}_{\bolds{2M}}$} &
\multicolumn{1}{c}{$\bolds{g}_{\bolds{3M}}$}&
\multicolumn{1}{c}{$\bolds{\hat{g}}_{\mathbf{1}\bolds{\mathit{PR}}}$} &
\multicolumn{1}{c}{$\bolds{\hat{g}}_{\mathbf{2}\bolds{\mathit{PR}}}$}&
\multicolumn{1}{c@{}}{$\bolds{\hat{g}}_{\mathbf{3}\bolds{\mathit{PR}}}$}\\
\hline
DE& Balanced & 792,210 & 97,376 & 44,113 & 459,930 & 120,989 & 124,922 &418,657
&142,987 & 268,279 \\
 & Unbalanced &1,129,602 &121,110 & 39,030 &937,376 &
116,803 & 75,548 &377,191 & 139,692 & 194,412 \\[2pt]
MD& Balanced & 792,210 & 696,375 & 66,040 &459,930 & 865,242 & 187,017 &418,657
&1,022,559 & 268,279 \\
 & Unbalanced &1,039,678 &858,915 &103,302 & 828,078 &
915,537 & 168,214 &409,823 & 1,157,267 & 229,034 \\
[2pt]
VA & Balanced &792,210 & 295,960 & 51,495 & 459,930 &367,728 & 145,828 &418,657
& 434,588 & 268,279 \\
& Unbalanced &863,313 &277,090 & 77,930 & 656,777 &
332,909 & 142,517 &485,759 &453,886 & 310,056 \\
[2pt]
WV& Balanced & 792,210 &106,701 & 71,651 & 459,930 & 132,575 & 202,908 &418,657
& 156,680 & 268,279 \\
 & Unbalanced& 839,191&84,658 & 67,832 &636,503 &100,218 &
144,901 &497,472 &135,006 & 322,621 \\
[2pt]
NC& Balanced & 792,210 & 125,904 & 43,600 & 459,930 & 156,435 & 123,472
&418,657 & 184,878 & 268,279 \\
 &Unbalanced& 583,606&60,730 & 43,268 & 447,884 & 85,908 &
98,855 &639,733 & 139,644 & 477,672 \\
[2pt]
SC & Balanced &792,210 & 110,835 & 2,312 & 459,930 & 137,711 & 6,548 &418,657 &
162,750 & 268,279 \\
 &Unbalanced&1,077,706& 126,061& 1,807 & 872,021 & 125,390
& 3,518 & 395,558&152,055 & 213,886 \\
[2pt]
GA& Balanced & 792,210 & 91,351 & 218,486 & 459,930 & 113,503 &618,727 &418,657
& 134,140 & 268,279 \\
 &Unbalanced& 716,079&57,972 & 235,098 & 540,917 & 74,935
&508,347 &562,280 & 108,923 & 392,592 \\
[2pt]
FL& Balanced & 792,210 & 134,749 & 144,408 & 459,930 & 167,426 & 408,947
&418,657 & 197,866 & 268,279 \\
 &Unbalanced& 605,744& 73,260 &125,413 & 462,990 &103,414
&293,078 & 626,452& 166,675 & 462,923 \\
[2pt]
AL & Balanced &792,210 & 94,683 & 163,838 & 459,930 & 117,643 & 463,970 &418,657
& 139,033 & 268,279 \\
 &Unbalanced& 774,894& 66,728 & 156,347 & 585,124 & 82,616
& 342,043 & 530,290& 115,370 & 357,952 \\
[3pt]
KY& Balanced &792,210 & 128,846 & 11,885 & 459,930 & 160,090 & 33,657 &418,657
& 189,198 & 268,279 \\
 &Unbalanced& 776,906&93,056 &13,089 & 586,680 &115,699 &
26,677 & 529,228& 162,238 & 356,806 \\
[3pt]
TN& Balanced &792,210 &136,323 & 18,482 & 459,930 & 169,381 & 52,340 &418,657 &
200,177 & 268,279 \\
 &Unbalanced& 769,515& 97,600 & 16,381 & 580,981 &122,136
& 37,950 & 533,140& 172,245 & 361,029 \\
[3pt]
MS& Balanced & 792,210 & 269,811 & 68,948 & 459,930 & 335,238 & 195,254
&418,657 & 396,191 & 268,279 \\
 &Unbalanced&1,060,717& 318,630 & 40,995 & 852,009 &
327,641 & 98,442 & 401,844& 405,527 & 220,560 \\
[3pt]
AR& Balanced & 792,210 & 239,946 & 28,598 & 459,930 & 298,131 & 80,987 &418,657
& 352,337 & 268,279 \\
 &Unbalanced& 918,002&229,443 & 18,346 & 705,154 & 260,944
& 45,703 & 460,413& 342,328 & 282,932 \\
[3pt]
LA & Balanced & 792,210 & 92,944 & 104,354 & 459,930 & 115,483 & 295,519
&418,657 & 136,480 & 268,279 \\
&Unbalanced& 879,552& 79,807 & 86,164 & 670,775 & 91,992
& 191,249 & 478,058& 121,522 &301,805 \\
[3pt]
OK & Balanced & 792,210 & 118,713 &164,196 & 459,930 & 147,500 & 464,983
&418,657 &174,318 & 268,279 \\
&Unbalanced&1,014,326& 130,894 & 113,081 &800,405 &
138,730& 255,182 & 419,730 & 174,242& 239,566 \\
\hline
\end{tabular*}\vspace*{3pt}
\end{table*}

In Table \ref{tab3.2} we include various components of the uncertainty measures
for the Prasad--Rao estimated MSE, the posterior variance of the HB
estimates and Morris' approximation to the HB moments. From these
components we can get the relevant overall uncertainty measure for the
EBLUPs, the HB estimates and Morris' approximation of EB estimates. We
note that in the balanced case the relative reduction in the
Prasad--Rao estimated MSE over the sampling variance (the measure of
uncertainty for the direct estimates) ranges between 21 and 62 percent.
These numbers clearly show substantial gain
in the accuracy of the model-based estimates. In the unbalanced setup,
these relative reductions range between $-23$ and 80 percent; only two
states, NC and FL, have negative improvement, which is somewhat
surprising. However, these two states being direct-use states in the
CPS, perhaps they enjoy large sample size to produce relatively
accurate direct estimates. Also, for these states, the $g_3$ term is
relatively big resulting in a large estimated MSE of the EBLUP. The
corresponding improvement numbers for the HB estimates are between 26
and 64 percent in the unbalanced case, and 28 and 58 percent in the
balanced case. For Morris' approximation, these numbers are between 21
and 69 percent in the unbalanced case, and 30 and 72 percent in the
balanced case.

In Table \ref{tab3.2} we present the decomposition of the
uncertainty\vadjust{\goodbreak}
corresponding to the three sources: uncertainty due to estimation of
unknown small area
mean, uncertainty due to estimation of the regression coefficients and
uncertainty due to unknown variance components. We consider the mean
squared error of an EBLUP (or EB predictor), the posterior variance and
its approximation due to Morris (\citeyear{Mor83N2}) for both the balanced
and an unbalanced setup. From this table we find that for each method
of estimation and each setup, all the three components of uncertainty
contribute substantially toward the overall measure of uncertainty for
most small areas. Thus it is important to account for the uncertainty
in estimating the regression coefficients and the variance components
in deriving a reliable overall measure of uncertainty associated with
the model-based small area
estimates.

In this example in the balanced Fay--Herriot setup the estimate of $A$
obtained by Prasad--Rao or Morris' method is $16\mbox{,}1617$, which is
substantially smaller than the HB estimate given by $1\mbox{,}735\mbox{,}616$. The
latter estimate is more than ten times the frequentist estimate and it
results from a very long tail of the posterior distribution of $A$.
This larger HB estimate of $A$ results in a substantially bigger value
of the first component (the $g_1$ term) of the Bayesian measure than
the corresponding component in the other measures. In fact the
frequentist estimate of $A$ is so small that, contrary to our
expectation, for some areas the estimate of the $g_1$ term is not the
dominant term in the estimated mean squared error (see the columns for
Morris' approximation and Prasad--Rao estimates).

We notice that the picture does not change substantially when we
consider the unbalanced setup. Here again, the posterior density of $A$
has a long tail resulting in a posterior mean of $2\mbox{,}063\mbox{,}419$. The
Prasad--Rao estimate is again far too small, only $192\mbox{,}527$, and
Morris's estimate is in between, which is $515\mbox{,}969$, much larger than
the Prasad--Rao estimate but much smaller than the HB estimate.

We reiterate that all three components contribute substantially toward
the overall measure of uncertainty. In particular, from the seventh and
the eighth columns of Table \ref{tab3.2}, we note that the third term (the~$g_3$
term) is bigger than the second term (the~$g_2$ term) in 14 of the 30 rows.
From the last two columns of the table, we note that the third term is
bigger than the second term in 22 of the 30 rows. All these indicate that
ignoring this component in the frequentist estimate of MSE or Morris'
estimate will result in a severe underestimation. It is particularly so
for the Prasad--Rao frequentist MSE since the first term ($g_1$ term)
is also adjusted for bias by adding the $g_3$ term. Incidentally, the
HB measure
of uncertainty automatically accounts for all sources of uncertainty.

We conclude this section noting that here and in the previous section
we assumed the sampling variances $V_i$'s are known. This assumption
was necessary to avoid the identifiability problem. If additional
independent estimates (independent of $Y_i$'s) of $V_i$'s are
available, and $V_i$'s
depend on a finite number of parameters, then the previous results can
be extended to develop model-based small area estimates of the means
and their measures of uncertainty. It can be done for both the EBLUP
and HB approaches. This is essentially similar to the unit-level model
considered briefly in Section \ref{sec6}. However, if $V_i$'s cannot
be assumed
to depend on a finite number of parameters, then the mean squared
approximation results presented here do not hold. In this scenario Wang
and Fuller (\citeyear{WanFul03}) assumed that independent ${\hat V}_i,
i=1,\ldots, m$,
are available which are unbiased for $V_i$. Assuming independent
chi-squared distributions of these estimates, they derived MSE
approximation of the EBLUP of $\theta_i$. Their approximation is valid
provided both $m$ and $d,$ the minimum of the degrees of freedom of the
chi-squared distribution, are large. Their approximation to the MSE is
accurate only to the order of $d^{-3/2}$. We refer to this article for
details. Another related paper in this setup is by Rivest and Vandal
(\citeyear{RivVan04}).

\section{Extensions}\label{sec4}
The Fay--Herriot (\citeyear{FayHer79}) model discussed in the previous
section can be
extended in different directions. First,
instead of $\mathbf{y}|\bth\sim N(\bth,\bG)$, where $\bG
=\operatorname{Diag}(V_1,\ldots
, V_m)$, one can begin with $\mathbf{y}|\bth\sim N(\bth
,\bV)$,
where $\bV$ is a known positive definite matrix which is not
necessarily diagonal. The full model is thus
%
%
\begin{equation}
\mathbf{y}|\bth\sim N(\bth,\bV),\quad \bth\sim N(\bX\bbe,
A\bI_m).
\label{e4p1}
\end{equation}
Datta et al. (\citeyear{Dat92}) considered this model in the context of
adjustment
of census undercounts. It is easy to check for
$A(>0)$ known, the BLUP (or the HB predictor with a flat prior for
$\bbe
$) is given by
%
%
\begin{equation}
\tilde{\bth}{}^B=(\bI_m-\bB)\mathbf{y}+\bB\bX\tilde
{\bbe}(A),
\label{e4p2}
\end{equation}
where $\bB=(\bV+A\bI_m)^{-1}\bV$ and $\tilde{\bbe}(A)=[\bX^T(\bV
+A\bI
_m)^{-1}\bX]^{-1}\bX^T(\bV+A\bI_m)^{-1}\mathbf{y}$. With $A$
unknown, one can opt either for estimation of $A$ from the marginal
distribution of $\mathbf{y}$, namely, $N(\bX\bbe, \bV
+A\bI_m)$
or put a flat prior for $A$, that is, $\pi(\bbe,A)=1$. Datta et al.
(\citeyear{Datetal92}) tried both methods in the context of adjustment of
census counts based on 1988 Missouri Dress Rehearsal data, but found
very little difference in the estimation of $\bth$.

The work of Datta et al. (\citeyear{Datetal92}) is based on modeling
the adjustment
factors related to census counts. To be specific, let $T_i$
denote the true count and the~$C_i$ the census count for the $i$th
small area. Then Cressie (\citeyear{Cre89}) and Isaki, Huang and Tsay
(\citeyear{IsaHuaTsa91})
proposed modeling $\theta_i=T_i/C_i$ $(i=1,\ldots, m).$

Direct estimates of these adjustment factors are usually obtained from
a postenumeration survey\break (PES) conducted by the Bureau of the
Census. In 1990,\vadjust{\goodbreak} the Bureau of the Census produced PES estimates of the
adjustment factors for 1,392 subdivisions (poststrata) of
the total population. The PES sample contained approximately $377\mbox{,}000$
persons in roughly 5,200 census blocks. However, prior to
the 1990 census, the Census Bureau had a trial run for several test
sites in Missouri to obtain direct estimates of these adjustment
factors based on (purported) complete enumeration and PES. Datta et al.
(\citeyear{Datetal92}) conducted an evaluation of this so-called
Census Dress
Rehearsal Data using the method described earlier in this section.

The HB and the EB estimators of $\bth$ based on (\ref{e4p2}) are given
respectively by
%
%
\begin{eqnarray}
\qquad\hat{\bth}{}^{\mathit{HB}} &=& [\bI_m -E(\bB|\mathbf{y})]\mathbf{y}+ E[\bB\bX\tilde{\bbe}(A)|\mathbf{y}],
\label{e4p3}\\
\hat{\bth}{}^{\mathit{EB}} &=& [\bI_m -\hat{\bB}]\mathbf{y}+ \hat
{\bB}\bX\tilde{\bbe
}(\hat{A}).
\label{e4p4}
\end{eqnarray}
The posterior variance $V(\bth|\mathbf{y})$, as before, is
given by
\begin{eqnarray*}
V(\bth|\mathbf{y}) &=& \bV\{\bI_m-E(\bB|\mathbf{y})\} \\
&&{}+ E[\bB\bX\{\bX^T(\bV+A\bI
_m)^{-1}\bX\}^{-1}\bX^T\bB^T|\mathbf{y}]\\
&&{} +\operatorname{Var}[\bB\{\mathbf{y}-\bX\tilde{\bbe}(A)\}
|\mathbf{y}] .
\end{eqnarray*}
This was found numerically very similar to the \textit{plug-in estimate}
of the
second-order approximate MSE given by
\begin{eqnarray*}
&& E[\{\bth-\hat{\bth}{}^{\mathit{EB}}\}\{\bth-\hat{\bth}{}^{\mathit{EB}}\}^T]\\
&&\quad\approx
\bV(\bI
_m-\bB) +
\bB\bX\{\bX^T(\bV+A\bI_m)^{-1}\}\bX^T\bB^T\\
&&\qquad{} +2\bV\bK^3\bV[\operatorname{tr}(\bV^{-2})]^{-1},
\end{eqnarray*}
where
\begin{eqnarray*}
\bK&=&(\bV+A\bI_m)^{-1} \\
&&{}-(\bV+A\bI_m)^{-1}\bX\{\bX^T(\bV+A\bI
_m)^{-1}\bX\}
^{-1}\\
&&{}\hspace*{11pt}\cdot\bX^T
(\bV+ A\bI_m)^{-1}.
\end{eqnarray*}
The study of Datta et al. (\citeyear{Datetal92}) revealed that for
every poststratum,
the EB (or EBLUP) and HB estimators
of the adjustment factors outperformed the direct estimators.

There is also a multivariate extension of the Fay--Herriot (\citeyear
{FayHer79}) model
considered in Datta, Fay and Ghosh~(\citeyear{DatFayGho91}). Now the data
consist of $\mathbf{y}_1,\mathbf{y}_2,\ldots
,\mathbf{y}_m$, where each $\mathbf{y}_i$ is
$r$-dimensional. Bivariate and trivariate versions of the model were
used in Datta, Fay and Ghosh (\citeyear{DatFayGho91}), and later in Datta et
al. (\citeyear{autokey13}) to
estimate median income of
four-person families for the $50$ states and the District of Columbia.
They considered the random effects model
%
%
\begin{equation}
\mathbf{y}_i=\bX_i\bbe+\bu_i+\be_i,\quad i=1,\ldots, m,
\label{e4p5}
\end{equation}
where $\bu_i\stackrel{\mathit{i.i.d.}}\sim N(\bdzero,\bA)$ and $\be_i\stackrel
{\mathit{ind}}
\sim N(\bdzero,\bV_i)$, the $\bu_i$ and the $\be_i$ being mutually\vadjust{\goodbreak}
independent,
and the $V_i(>0)$ are known. Alternatively,\vspace*{-1pt} in a Bayesian fra\-mework,
writing\vspace*{-1pt} $\bth_i=\bX_i\bbe+\bu_i$ $(i=1,\ldots, m)$, $\mathbf{y}_i|\break\bth_i
\stackrel{\mathit{ind}}\sim N(\bth_i,\bV_i)$ and $\bth_i\stackrel
{\mathit{ind}}\sim N(\bX
_i\bbe,\bA)$. Both EB (or EBLUP) and HB estimators of
the $\bth_i$ were found. These estimators were shown to outperform the
direct estimators with respect to their precision measures.

\section{Confidence Intervals in Small Area Estimation}\label{sec5}
Morris (\citeyear{Mor83N2}) noted that although Stein's shrinkage
estimators were
widely used for point estimation, a lack of the availability
of estimated uncertainty with these estimators delayed development of
reliable confidence intervals. An early attempt to construct
EB confidence intervals is due to Cox (\citeyear{Cox75}). In the small area
estimation terminology, he developed approximate confidence intervals
that are accurate to the order of $O(m^{-1})$ for an individual small
area mean $\theta_i$ for the balanced Fay--Herriot model without any
covariate. Again in the small area estimation terminology, Morris
(\citeyear{Mor83N1,Mor83N2}) was the first to consider confidence
intervals for small area
means for the Fay--Herriot model with covariates. He considered both
the balanced and the unbalanced sampling variance cases. His method
consists essentially in finding an HB confidence interval for $\theta
_i$, approximating (using Laplace approximations to integrals)
this interval with estimates of the hyperparameters only at the last
stage. He constructed these intervals using normal percentile
points and provided a heuristic justification of these naive EB
intervals. Later Laird and Louis (\citeyear{LaiLou87}) proposed EB bootstrap
confidence intervals in the spirit of
Morris (\citeyear{Mor83N1,Mor83N2}), while Carlin and Gelfand
(\citeyear
{CarGel90}), following a
suggestion of Efron, proposed calibrating the naive EB confidence intervals.
Indeed, in small area estimation setup, both for unit-level and
area-level data, Prasad and Rao (\citeyear{PraRao90}) also suggested
approximate
confidence intervals for small area means. They based their intervals
on normal percentile points and used their second-order unbiased estimator
of the MSE of the EBLUP. As in Morris (\citeyear{Mor83N1,Mor83N2}),
Prasad--Rao intervals
also have a coverage error to the order of $O(m^{-1})$.

For the case when $V_1=\ldots,V_m=V$ and $A$ are both known, a flat prior
for $\bbe$ will result in a $100(1-\alpha)\%$ confidence interval of the
form $(1-B)y_i+B\mathbf{x}_i^T\hat{\bbe}\pm z_{\alpha
/2}V^{1/2}(1-B+Bm^{-1})^{1/2}$,
where we may recall that $\hat{\bbe}=(\bX^T\bX)^{-1}\bX^T\mathbf{y}$, the least
squares estimator of $\bbe$. This result\vadjust{\goodbreak} follows immediately from Lindley
and Smith (\citeyear{Li72}). A naive EB confidence interval is given by
$(1-B)y_i+B\mathbf{x}_i^T\hat{\bbe}\pm z_{\alpha
/2}V^{1/2}(1-B)^{1/2}$, which
does not take into account uncertainty due to estimation of $\bbe$.
Accordingly, the coverage~pro\-bability will fall short of the target
under the
said hierarchical model. The Type III bootstrap approach of Laird and Louis
(\citeyear{LaiLou87}) provides a confidence interval identical to the
hierarchical Bayesian
approach, where the bootstrap samples $\mathbf{y}$ are drawn
from the
$N(\bX\hat{\bbe},(V+A)\bI_m)$ pdf. The same confidence interval is also
arrived at by the conditional~ap\-proach of Hill (\citeyear{Hil90}).
Hill's approach
consists of~fin\-ding the conditional distribution of $\theta_i-((1-B)y_i+
B\mathbf{x}_i^T\hat{\bbe})$ given the ancillary statistic
$U_i=y_i-\mathbf{x}_i^T\hat{\bbe}$. Also, it is pointed
out by Laird and Louis
(Theorem~2.1, page~743) that the Type III bootstrap can never match a~hyperprior solution when $A$ is unknown.

For the balanced Fay--Herriot model, Datta et al. (\citeyear
{Datetal02}) developed an
expansion for the coverage pro\-bability
of confidence intervals derived by Morris\break (\citeyear{Mor83N1,Mor83N2})
and Prasad and Rao
(\citeyear{PraRao90}). Based on this expansion they perturbed the
endpoints of the confidence interval to achieve asymptotic coverage accurate
to the order of $o(m^{-1})$. Also, following
the framework of Hill (\citeyear{Hil90}), Datta et al. (\citeyear
{Datetal02}) studied conditional coverage
probabilities of such intervals even for unknown $A$ by conditioning on
a suitable ancillary statistic. They obtained an expansion of the
conditional coverage probability as well and used the expansion to better
calibrate the interval. For some $d>p$ let $\hat{B}(S)\equiv\hat
{B}_d(S) =(m-d)\operatorname{min}\{V/S, (m-p)^{-1}\}$. Assuming
${\operatorname{max}}_{1\le i\le m}h_{ii}=O(m^{-1}),$ where $h_{ii}$ is defined
in Theorem \ref{th3},
they had for any fixed $t>0$ the following expansion.

\begin{theorem}\label{th4}
\begin{eqnarray*}
&&\hspace*{-5pt} P\bigl[\theta_1 \in\bigl(1-\hat B(S)\bigr)Y_1 + \hat B(S) \mathbf{x}_1^T
\hat{\bbe} \pm t V
\bigl(1-\hat B(S)\bigr)^{1/2}\bigr]
\\
&&\quad= 2\Phi(t)-1\\
&&\qquad{} - t\phi(t)\biggl[\frac{(1+t^2)B^2}{2m(1-B)^2} + \frac{B}{1-B}
\biggl\{h_{11} + \frac{5-d}{m}\biggr\}\biggr]
\\
&&\qquad{}+ O(m^{-3/2}).
\end{eqnarray*}
\end{theorem}

Let $z_{\alpha/2}$ denote the upper $\alpha/2$ point of $N(0,1)$
distribution. Taking $t=z_{\alpha/2}$ will result in an under\-estimation
in the nominal coverage $1-\alpha$. If we take
\[
t^*=z_{\alpha/2}\biggl[1 + \frac{(1+z_{\alpha/2}^2)\hat{B}^2}{4m(1-\hat
{B})^2} + \frac{(5-d+mh_{11})\hat{B}}{2m(1-\hat{B})}\biggr],
\]
it follows that the interval $(1-\hat B(S))Y_1 + \hat B(S) \mathbf{x}_1^T \hat
{\bbe} \pm t^* V (1-\hat B(S))^{1/2}$
has coverage probability\vadjust{\goodbreak} equal to $1-\alpha$ up to $O(m^{-3/2})$ error
terms. Although this theorem is presented in
the context of EB intervals, Datta et al. (\citeyear{Datetal02}) also discussed
expansion of coverage probabilities of intervals that
are created through the HB argument of Morris (\citeyear{Mor83N1}).

Extending the argument of Hill (\citeyear{Hil90}), Datta et al.
(\citeyear{Datetal02}) also
obtained an expansion of the coverage probability of an
EB confidence interval of $\theta_1$ by conditioning on an ancillary
statistic $U=(Y_1-\mathbf{x}_1^T\hat{\bbe})\sqrt
(m-p)/\sqrt{S}$. They
proved the following theorem.

\begin{theorem}\label{th5}
\begin{eqnarray*}
&&P\bigl[\theta_1 \in\bigl(1-\hat B (S)\bigr)Y_1+\hat B(S) \mathbf{x}_1^T
\hat{\bbe} \\
&&{}\hspace*{50pt}\pm t V
\bigl(1-\hat B(S)\bigr)^{1/2}|U\bigr]
\\
&&\quad= 2\Phi(t) -1\\
&&\qquad{} - t\phi(t) \biggl[\frac{(1+t^2)B^2}{2m(1-B)^2} +
\frac{(2U^2+3-d)B}{m(1-B)}
%
%
\\
&&\hspace*{166pt}{}+ \frac{B h_{11}} {1-B}\biggr]\\
&&\qquad{}+ O_p(m^{-3/2}).
\end{eqnarray*}
\end{theorem}

The bias corrected confidence intervals for $\theta_1$ are obtained as
before with appropriate changes.

Datta et al. (\citeyear{Datetal02}) performed a simulation study to
evaluate the
performance of the approximate confidence intervals given in
the two theorems above. In these simulations they used a simple setup
with $m=30$ small areas with no covariates. Since the coverage
probability does not depend on $\bbe$, it was taken as zero in
generating the samples. Also, the coverage probability depends only on $B$,
so without any loss of generality $V$ was taken to be $1$. These
authors considered various values of $B$ in the range $0.025$ to $0.975$.
They computed both conditional and unconditional coverage probabilities
as discussed in the theorems given above. They found little
qualitative difference in performance between the unconditional and
conditional coverage probabilities. They also noted that while the
extent of
underestimation of the coverage probabilities with $t=z_{\alpha/2}$
from the nominal level $\alpha$ was small for small $B$, the underestimation
was severe for $B$ in the upper half. On the other hand, the adjusted
intervals appeared to be too large resulting in overestimation of the
coverage probabilities. This overestimation is due to an overestimation
of the mean squared error of the EB estimator of $\theta_1$. Incidentally,
Lahiri and Rao (\citeyear{LahRao95}) also noted similar overestimation
of the MSE when~$B$ approaches 1, that is, when $A/V$ approaches $0$.\vadjust{\goodbreak}

Smith (\citeyear{Smi}) in his unpublished Ph.D. dissertation developed
EB confidence
intervals for the $i$th small area mean $\theta_i$ for the more
practical case of~un\-balanced Fay--Herriot model in (\ref{e3p1}).
Associated with
the EB or EBLUP $\hat{\theta}_i^{\mathit{EB}}$ of $\theta_i$, let $s_i^2$ denote
some esti\-mated measure of uncertainty. Note that~$s_i^2$ could~be a second-order unbiased estimator of
 the MSE of~$\hat{\theta
}_i^{\mathit{EB}}$ as in (\ref{mse2}) or something similar. For some estimator
$\hat{A}$
of $A$, Smith (\citeyear{Smi}) defined \mbox{$s_i^2=h_i^2(\hat{A})+c_i$}, where
$h_i^2(A)=g_{1i}(A)+g_{2i}(A)$. The term~$c_i$ is an $O_p(m^{-1})$
order term, that may depend on~$\hat{A}$ and the data $\bY$, and may be
related to $g_{3i}$ term in (\ref{e3p14}) and bias term
of $\hat{A}$. There are
many possible choices corresponding to various MSE estimates. Rao
(\citeyear{Rao01}) proposed a number of area-specific estimators of the
MSE of the
EBLUP, and
they can be included by proper choice of $c_i$. Alternatively, in the
HB setup,
$c_i$ may include $(y_i-\mathbf{x}_i^T\hat{\bbe})^2
\operatorname{Var}(B_i(A)|\mathbf{y})$, which is
an approximation to the last term in the
posterior variance in (\ref{e3p15}). This general choice enabled Smith
to study approximate coverage probabilities of confidence intervals constructed
in Morris (\citeyear{Mor83N1,Mor83N2}) by using EB and HB methods.
Corresponding to
$c_i$, let the parametric function $c_i^*(A)$ be such that
$c_i-c_i^*(A) =o_p(m^{-1})$. Also, define $q_i(A)=B_i^2(A)b_{\hat{A}}(A)
+c_i^*(A)-2g_{3i}(A)$, where $b_{\hat{A}}(A)$ is the asymptotic bias of
$\hat{A}$. With the above notation we now state Theorem 1.7.1 of
Smith (\citeyear{Smi}) below.

\begin{theorem}\label{th6}
For any $z>0$,
\begin{eqnarray*}
& & P[\hat{\theta}_i^{\mathit{EB}} -z s_i \le\theta_i \le\hat{\theta
}_i^{\mathit{EB}} +
z s_i] \\
&&\quad= 2\Phi(z) -1 + z\phi(z)\frac{q_i(A)}{h_i^2(A)}
\\
&&\qquad{}- \frac{(z^3+z)\phi(z)g_{3i}(A)}{4h_i^4(A)}\frac{D_i^2}{(A+D_i)}+ o(m^{-1}).
\end{eqnarray*}
\end{theorem}

Note that the leading term in the above expansion is the nominal
coverage probability. The first-order error term in
this expansion is of order $O(m^{-1})$. From this expansion it follows
that as in Theorem~\ref{th4} we can perturb the cut-off point
$z$ in order to achieve the nominal coverage probability to the order
$o(m^{-1})$. Another point to note is that since the $O(m^{-1})$
term~$c_i$ (or equivalently, $c_i^*$) was not completely specified, for
any given $z$ we can choose $c_i^*(A)$ (depending on $z$ and $A$)
to make the $O(m^{-1})$ term in the expansion of the coverage
probability disappear. In particular, the choice $c_i=c_i^*(\hat{A})$ with
\[
c_i^*(A)=2g_{3i}(A)-B_i(A)^2b_{\hat{A}}(A) +
\frac{(z^2+1)D_i}{4A}g_{3i}(A)\vadjust{\goodbreak}
\]
will give an EB confidence interval that matches the nominal coverage
probability to the order of $o(m^{-1})$.

In this section we have considered confidence intervals for individual
small area means, which is the current state of the literature. In the
early applications of small area estimation, practitioners were only
interested in point estimates (see, e.g., Fay and Herriot, \citeyear
{FayHer79}). Only
in the last twenty years or so, substantial development of the measures
of uncertainty of the model-based estimates of small area means has
taken place. Construction of appropriate confidence intervals for small
area means is still limited and is restricted only to individual means.
While in the EB
setup confidence sets for several population means have been
considered, this problem is not fully addressed yet in small area
estimation. In a recent article, Ganesh (\citeyear{Gan09}) has considered
simultaneous credible intervals in small area estimation. However,
calibrated confidence sets for multiple small area means in EB or EBLUP
approach have not been studied yet.

\section{Other Important Developments in~Small Area Estimation}\label{sec6}
We mentioned in the Introduction that both area-level and unit-level
data are available in small area estimation. In the previous
sections we have concentrated mostly on area-level models. In this
section we review some of the results for unit-level models.
For a unit-level model let $y_{ij}$ denote the value for the $j$th
unit in the $i$th small area, with $j=1,\ldots, N_i$, $i=1,\ldots,
m$, where
$N_i$ is the size of the finite population corresponding to the $i$th
small area. Let
$\gamma_i=N_i^{-1}\sum_{j=1}^{N_i} y_{ij}$ denote the finite population
mean for the $i$th small area. For notational
simplicity let $y_{ij}, j=1,\ldots, n_i, i=1,\ldots, m$ denote values of
the characteristic of the sampled units from
these $m$ small areas. Let the vector $\mathbf{y}(s)$ denote
all the sampled
values. A direct estimator of $\gamma_i$ based on the
$i$th area sample mean $\bar{Y}_{is}$ is usually less reliable due to
a small sample size $n_i$. To borrow strength from the
neighboring areas through shrinkage estimation the following model,
known as the nested-error regression
model, has been found very useful for unit-level data. The model is
given by
%
%
\begin{eqnarray}\label{e5}
&& Y_{ij}=\mathbf{x}_{ij}^T\bbe+v_i+e_{ij},
\nonumber
\\[-8pt]
\\[-8pt]
\nonumber
&&\quad j=1,\ldots, N_i,
i=1,\ldots, m,
\end{eqnarray}
where $\mathbf{x}_{ij}$ is a $p$-component vector of
auxiliary variables,
$v_i$ and $e_{ij}$ are independently distributed with\vspace*{-2pt}
$v_i \stackrel{\mathit{i.i.d.}}\sim N(0,\sigma^2_v)$ and $e_{ij} \stackrel
{\mathit{ind}}\sim
N(0,\sigma^2_e)$, $j=1,\ldots, N_i,  i=1,\ldots, m$.
We denote the observations for the samp\-led units in the $i$th small
area by
$\bY_i^{(1)}=(Y_{i1},\ldots,\break  Y_{in_i})^T$.\vspace*{-1pt}
Similarly, $\bY_i^{(2)}$ is used to denote
the vector of observations corresponding to the unsampled units in the
$i$th small area. Battese, Harter and Fuller (\citeyear{BatHarFul88}) and
Prasad and Rao (\citeyear{PraRao90}) used this model to develop EBLUP
estimate of
finite population mean $\gamma_i$. They approximated
$\gamma_i$ for large $N_i$ by $\theta_i=\bar{\bX}_i^T\bbe+v_i$ and
used the predictor of $\theta_i$ to estimate $\gamma_i$.
Here $\bar{\bX}_i=N_i^{-1}\sum_{j=1}^{N_i} \mathbf{x}_{ij}$ is the known mean
vector of the auxiliary variables.

Let $\bY^{(1)}$ be obtained by stacking the vectors $\bY_i^{(1)}$ for
all the $m$ small areas. Similarly, denote by
$\bX^{(1)}$ the matrix of $p$ columns obtained by stacking\break the~$\mathbf{x}
_{ij}$'s corresponding to the sampled units. We also
denote the variance of $\bY^{(1)}$ by $\bSi_{11}$. From Prasad and Rao
(\citeyear{PraRao90}) the BLUP of $\theta_i$ is obtained as
\begin{equation}
\tilde{\theta}_i\bigl(\bolds{\psi},\bY^{(1)}\bigr)= \bar{\bX}
_i^T\tilde{\bbe} +
\delta_i(\bar{Y}_{is} - \bar{\mathbf{x}}_{is}^T\tilde{\bbe}),
\label{e12}
\end{equation}
where $\bolds{\psi}=(\sigma_v^2,\sigma_e^2)$, and
\begin{equation}
\tilde{\bbe} = \bigl(\bX^{(1)T}\bSi_{11}^{-1}\bX^{(1)}\bigr)^{-1}\bX^{(1)T}
\bSi_{11}^{-1}\bY^{(1)}
\label{e11}
\end{equation}
is the generalized least squares estimator of $\bbe$. Here $\delta
_i=\sigma_v^2(\sigma_v^2+\sigma_e^2n_i^{-1})^{-1}$
is the shrinkage coefficient which shrinks the direct estimator $\bar
{Y}_{is}$ of $\gamma_i$ (or $\theta_i$) toward
a regression surface.

Under the superpopulation model given by (\ref{e5}), from
Prasad and Rao (\citeyear{PraRao90}) and Datta and Ghosh (\citeyear
{DatGho91N2}) one can show that the
BLUP of the finite
population mean $\gamma_i$ under the nested error regression model
is given by
\begin{equation}
\qquad\tilde{\gamma}_i\bigl(\bolds{\psi},\bY^{(1)}\bigr)= f_i \bar{Y}_{is}
+(1-f_i)\tilde{\theta}_{i(u)}\bigl(\bolds{\psi},\bY^{(1)}\bigr),
\label{e14}
\end{equation}
where
$f_i=n_i/N_i$, $\tilde{\theta}_{i(u)}(\bolds{\psi},\bY
^{(1)})$ is given by
(\ref{e12}), with $\bar{\bX}_i$ replaced by $\bar{\mathbf{x}}_{i(u)}$, the
mean of $\mathbf{x}_{ij}$'s for the $N_i-n_i$ unsampled
units from the
$i$th area.
The BLUP of the small area mean $\gamma_i$ usually depends on
variance components, which in practice will be unknown. Estimates of variance
components $\bolds{\psi}$ are plugged in to the BLUP to
obtain EBLUP
estimates. The variance components are estimated from
the marginal distribution (by integrating out $v_i$'s) of the data,
$\bY^{(1)}$.

While Datta and Lahiri (\citeyear{DatLah00}) suggested ML and REML
estimation of the
variance components, Pra\-sad and Rao (\citeyear{PraRao90})
used ANOVA methods to obtain unbiased estimators for variance
components in the nested error regression model. Prasad and Rao
(\citeyear{PraRao90})
first obtained
$\hat{e}_{ij}$, $\hat{u}_{ij}$, $j=1,\ldots, n_i$, $i=1,\ldots, m,$ whe\-re
$\{\hat{e}_{ij}, j=1,\ldots, n_i, i=1,\ldots,m \}$\vadjust{\goodbreak} are the residuals
from the ordinary least squares regression of $Y_{ij}-\bar{Y}_{is}$
on $\{{\mathbf{x}}_{ij}-\bar{\mathbf{x}}_{is}\}$
and $\hat{u}_{ij} $ are the
residuals from the ordinary least squares regression of $Y_{ij}$ on
$\mathbf{x}_{ij}$. Estimators
\begin{eqnarray}\label{e15}
\hat{\sigma}^2_e&=&(n-m-p^*)^{-1}\sum_{i=1}^m\sum_{j=1}^{n_i}\hat{e}_{ij}^2,\quad
\mbox{and}
\nonumber
\\[-8pt]
\\[-8pt]
\nonumber
\hat{\sigma}^2_v&=&n_*^{-1}\Biggl[\sum_{i=1}^m\sum_{j=1}^{n_i}\hat{u}_{ij}^2
-(n-p)\hat{\sigma}^2_e\Biggr]
\end{eqnarray}
are unbiased, where $n_*=n- \operatorname{tr}[(\bX^{(1)T}\bX^{(1)})^{-1}\cdot\break \sum_{i=1}^m
n_i^2\bar{\mathbf{x}}_{is}\bar{\mathbf{x}}_{is}^T]$, and $p^*$ is equal to
the number of linearly independent vectors in the set
$\{{\mathbf{x}}_{ij}-\bar{\mathbf{x}}_{is},
j=1,\ldots, n_i, i=1,\ldots, m\}$.

Second-order accurate approximations to MSE of the EBLUP of $\theta_i$
were developed by
Prasad and Rao (\citeyear{PraRao90}) and Datta and Lahiri (\citeyear
{DatLah00}). These authors showed
for the nested error regression
model the three terms in the approximation [cf. (\ref{e3p14})] are
\begin{eqnarray}\label{e25}
g_{1i}(\bolds{\psi})&=& (1 -\delta_i)\sigma_v^2,
\nonumber
\\
\qquad\quad g_{2i}(\bolds{\psi}) &=& (\bar{\bX}_i - \delta_i \bar
{\mathbf{x}}_{is})^T
\bigl(\bX^{(1)T} \bSi_{11}^{-1}(\bolds{\psi}) \bX
^{(1)}\bigr)^{-1}\\
&&{}\cdot (\bar{\bX}_i -
\delta_i \bar{\mathbf{x}}_{is}),\nonumber
\\
\label{e26}g_{3i}(\bolds{\psi})&=&
n_i^{-2}(\sigma_v^2 +\sigma_e^2/n_i)^{-3}
\nonumber
\\[-8pt]
\\[-8pt]
\nonumber
&&{}\cdot \operatorname{var}
(\sigma_v^2\hat{\sigma}_e^2 - \sigma_e^2\hat{\sigma}_v^2).
\end{eqnarray}
For an estimator $\hat{\bolds{\psi}}$ of $\bolds{\psi}$, from Prasad and Rao
(\citeyear{PraRao90})
and Datta and Lahiri (\citeyear{DatLah00})
a second-order unbiased estimator of the MSE of the EBLUP of $\theta_i$
is given by
\begin{eqnarray}\label{e33}
\mathit{mse}(\tilde{\theta}_i(\hat{\bolds{\psi}}))&=&g_{1i}(\hat
{\bolds{\psi}})+ g_{2i}(\hat
{\bolds{\psi}})+ 2g_{3i}(\hat{\bolds{\psi}})
\nonumber
\\[-8pt]
\\[-8pt]
\nonumber
&&{}-\mathbf{b}^T(\hat{\bolds{\psi}};\hat
{\bolds{\psi}})
\nabla g_{1i}(\hat{\bolds{\psi}}),
\end{eqnarray}
where $\mathbf{b}^T(\hat{\bolds{\psi}};{\bolds{\psi}})$ is the asymptotic bias of
$\hat{\bolds{\psi}
}$, and $\nabla g_{1i}({\bolds{\psi}})$ is the gradient
vector of $g_{1i}({\bolds{\psi}})$. For estimators of
variance components with
asymptotic bias of $o(m^{-1})$, the last term in
(\ref{e33}) drops out. This happens for the ANOVA estimators suggested
by Prasad and Rao (\citeyear{PraRao90}) and the REML estimators considered
by Datta and Lahiri (\citeyear{DatLah00}).

Estimation of the MSE of EBLUP outlined above and in Section \ref{sec3}
is based
on Taylor's expansion. Alternatively, a resampling-based
approach may be used to estimate the MSE. Laird and Louis (\citeyear{LaiLou87})
suggested a bootstrap measure of accuracy of the EB estimator for
the Fay--Herriot model. Subsequently, Butar and Lahiri (\citeyear
{ButLah03}) adopted
their approach in small area estimation.\vadjust{\goodbreak} Further references
to this lite\-rature may be found in Pfeffermann and Tiller\break (\citeyear
{PfeTil05}),
Lahiri (\citeyear{Lah03}) and Hall and Maiti (\citeyear{Ha07}). Jiang, Lahiri and Wan (\citeyear{JiaLahWan02})
proposed jackknife methods to estimate the MSE of the EBLUP.

Datta and Ghosh (\citeyear{DatGho91N2}) proposed a general HB model for
unit-level data
in small area estimation. Some earlier Bayesian
analysis for two-stage sampling in a simpler framework is due to Scott
and Smith (\citeyear{ScoSmi69}), with subsequent extension to the
multistage sampling by Malec and Sedransk (\citeyear{MalSed85}). Ba\-sed
on the
superpopulation approach to finite population sampling Datta
and Ghosh (\citeyear{DatGho91N2}) developed HB estimates of small area
means by
deriving certain predictive distributions. To that objective, they
considered the following HB model:
\begin{enumerate}[{(A)}]
\item[{(A)}] Conditional on $\bbe$, $\bla=(\lambda_1,\ldots
,\lambda
_t)^T$ and $r$, let
\[
\bY\sim N\bigl(\bX\bbe,r^{-1}\bigl(\bPsi+\bZ\bD(\bla)\bZ^T\bigr)\bigr),
\]
where $\bY$ is $N\times1$ vector of characteristics of all the $N$
units in the finite population, $\bX$ and $\bZ$ are $N\times p$ and
$N\times q$ matrices, respectively, for appropriate known $p$ and $q$.
\item[{(B)}] $\bbe, r$ and $\bla$ have a certain joint prior distribution.
\end{enumerate}

Stage (A) of the above model can be identified as a general mixed
linear model (cf. Datta and Ghosh, \citeyear{DatGho91N2}). To see
this, write
\beq
\bY=\bX\bbe+\bZ\mathbf{v}+\be,
\label{mixed}
\eeq
where $\be$ and $\mathbf{v}$ are mutually independent with
$\be\sim N(\bdzero,
r^{-1}\bPsi)$, and $\mathbf{v}\!\sim\!N(\bdzero, r^{-1}\bD
(\bla))$. Here
$\be$ is \mbox{$N\!\times\!1$}, and $\mathbf{v}$ is $q\times1$
vector of random
effects, $\bPsi$ is a~known positive definite matrix and $\bD(\bla)$ is
a $q\times q$ p.d. matrix which is known except for $\bla$.

In the context
of finite population $\bY$ is partitioned as $\bY^T=(\bY^{(1)T},\bY
^{(2)T})$, where $\bY^{(1)}$ corresponds to the sampled units
and $\bY^{(2)}$ corresponds to the unsampled units. Similarly, the
design matrices $\bX$ and~$\bZ$ are partitioned. To make inference
about certain functions of $\bY$, the Bayesian solution is obtained by
deriving the predictive distribution of~$\bY^{(2)}$ given
$\bY^{(1)}=\mathbf{y}^{(1)}$ (which is the posterior
distribution of~$\bY^{(2)}$). In small area estimation the vector of sampled units
$\bY^{(1)}$ is from $m$ small areas. If $\bY_i^{(1)}$ is the
$(n_i\times1)$ vector of sampled units from the $i$th small area,
then $\bY^{(1)T}=(\bY_1^{(1)T},\ldots, \bY_m^{(1)T})$. Similarly, the
vector $\bY^{(2)}$ corresponding to the unsampled units can be
partitioned. The finite population mean $\gamma_i$ from small area $i$
is a linear function of~$\bY^{(2)}$, and its predictive distribution
may be derived from the distribution of $\bY^{(2)}$. In particular,
based on a\vadjust{\goodbreak} quadratic loss function, the HB estimator is given
by the posterior mean of $\gamma_i$, and a measure of uncertainty is
given by the posterior variance of $\gamma_i$. While the
solution for the general HB model is presented in Datta and Ghosh
(\citeyear{DatGho91N2}), we now spell out below some of the details for
the nested error
regression model.

For the nested error regression model in (\ref{e5}), $t=1$, $r=\sigma_e^{-2}$
and $\lambda_1=\sigma_e^{2}/\sigma_v^{2}$. To complete the
HB model, Datta and Ghosh (\citeyear{DatGho91N2}) assigned independent prior
distribution on\vspace*{1pt} $\bbe$, $\sigma_e^2$ and $\sigma_v^2$. They put a~uniform
prior over $R^p$ for $\bbe$, and $\sigma_e^2\sim \mathit{IG}(a_0/2,\break g_0/2)$ and
$\sigma_v^2\sim \mathit{IG}(a_1/2,g_1/2)$, where $\mathit{IG}(\beta,\alpha)$ is
a~distribution whose pdf is proportional to $\exp(-\beta/\break x)
x^{-\alpha-1}$. Quantities $a_0, g_0, g_1$ are nonnegative and~$a_1$
is positive, and are chosen suitably small to reflect diffused prior
information on the variance components.

The HB estimates for any reasonably complex mo\-del do not admit any
closed-form expressions, and they are evaluated by numerical computations.
Required posterior moments can be found either by Gibbs sampling (cf.
Gelfand and Smith, \citeyear{GelSmi90}) or by numerical integration.
Using formulas
for iterated expectation and variance, Datta and Ghosh (\citeyear
{DatGho91N2}) have
shown that the posterior mean and the posterior variance can be computed
by evaluating several one-dimensional integrals with respect to the
posterior density of $\lambda_1$. In particular, the HB estimate of~$\gamma_i$ is
\[
\hat{\gamma}_i{}^{\mathit{HB}}=E\bigl[\tilde{\gamma}_i\bigl(\lambda_1,\mathbf{y}^{(1)}\bigr)|\mathbf{y}^{(1)}\bigr],
\]
where the expectation $E[\cdot|\mathbf{y}^{(1)}]$ is with
respect to the
posterior density of $\lambda_1$, and $\tilde{\gamma}_i(\lambda
_1,\mathbf{y}^{(1)})$
(with a slight abuse of notation) is the same as the expression
of~$\tilde{\gamma}_i(\bolds{\psi},\bY^{(1)})$ given in
(\ref{e14}). Note that
the above HB estimate of $\gamma_i$ is obtained by shrinking the direct
small area estimator $\bar{Y}_{is}$ to an estimated regression surface.
Similarly, the posterior variance of~$\gamma_i$ can also be computed by
numerical integration involving one-dimensional integrals. Alternatively,
the Gibbs sampling can also be implemented very easily for the present
model. Indeed Datta and Ghosh (\citeyear{DatGho91N2}) have shown that
the set of
complete conditional distributions are given by either multivariate normal
or inverse gamma distributions.

\section{Other Small Area Estimators}\label{sec7}
\subsection{Measurement Error Models}\label{sec7.1}

In our presentation of the unit-level model, we have assumed so far
that the covariates are measured without error. However, sometimes
it is not possible\vadjust{\goodbreak} to obtain exact measurements of these covariates.
For example, if in prediction of certain crop yield, the nitrogen
level in the soil is a covariate, this covariate needs to be determined
by analysis of soil sample. This will result in measurement error
of the covariate. For the nested error regression model with a single
covariate with measurement error Ghosh and Sinha (\citeyear{GhoSin07}),
Ghosh, Sinha and Kim
(\citeyear{GhoSinKim06}) and Torabi, Datta and Rao (\citeyear{TorDatRao09})
have considered estimation of small
area means. While Ghosh and Sinha (\citeyear{GhoSin07}) used a
functional measurement error
model, Ghosh, Sinha and Kim (\citeyear{GhoSinKim06}) and Torabi, Datta and Rao
(\citeyear{TorDatRao09}) considered a
structural measurement error model for estimation of small area means. Ghosh, Sinha and Kim (\citeyear{GhoSinKim06}) and Torabi, Datta and Rao (\citeyear
{TorDatRao09}) used the model given by
%
%
\begin{eqnarray}
\label{1}
&&y_{ij}=\beta_{0}+\beta_{1}x_{i}+v_{i}+e_{ij};
\nonumber
\\[-8pt]
\\[-8pt]
\nonumber
&&\quad j=1,\ldots ,n_{i};i=1,\ldots ,m,
\end{eqnarray}
where as before $y_{ij}$ is the response variable of the $j$th unit in
the $i$th area (or
stratum), $x_{i}$ is the unknown true area-specific covariate
associated with\vspace*{-1pt} $y_{ij}$. Further, $v_{i} \stackrel{\mathit{i.i.d.}}{\sim}
N(0,\sigma^{2}_{v})$ and independent of $e_{ij} \stackrel
{\mathit{i.i.d.}}{\sim}
N(0,\sigma^{2}_{e}).$ Under measurement errors,\vspace*{-2pt}
$X_{ij}(=x_{i}+u_{ij})$ are
observed, where $u_{ij} \stackrel{\mathit{i.i.d.}}{\sim} N(0,\sigma^{2}_{u}).$
They\vspace*{-2pt} also assumed that $x_{i} \stackrel{\mathit{i.i.d.}}{\sim}
N(\mu_{x},\sigma^{2}_{x})$. The vector of model parameters is given by
$\theta=(\beta_{0},\beta_{1},\mu_{x},\sigma^{2}_{x},\sigma
^{2}_{u},\sigma^{2}_{v},\sigma^{2}_{e})^{T},$
and $x_{i},v_{i},e_{ij}$ and $u_{ij}$ are assumed to be mutually
independent.

Based on the preceding model Ghosh, Sinha and Kim (\citeyear{GhoSinKim06})
obtained the EB
predictor of $\gamma_i$ by replacing the model parameters
by their estimates in the Bayes estimator of $\gamma_i$ based on the
conditional distribution of $Y_{ij}, j=n_i+1,\ldots, N_i$,
given $\theta$ and~$y_{ij}$, $j=1,\ldots, n_i$. Since $X_{ij}$'s are also
stochastic Tora\-bi, Datta and Rao (\citeyear{TorDatRao09}) instead first derived
the fully efficient
Bayes estimator of $\gamma_i$ based on the conditional distribution of
$Y_{ij}, j=n_i+1,\ldots, N_i$,
given~$\theta$, $y_{ij}, j=1,\ldots, n_i$, and $x_{ij}, j=1,\ldots, n_i$.
Finally, they obtained an EB estimate of $\gamma_i$ by replacing
$\theta$, the model parameters by their estimates as given in  Ghosh, Sinha and Kim (\citeyear{GhoSinKim06}).
Torabi, Datta and Rao (\citeyear{TorDatRao09}) employed
the jackknife
method to obtain an estimate of mean squared prediction error (MSPE) of
the EB predictor. For further details we refer to these two papers.

\subsection{Generalized Linear Models}\label{sec7.2}

Until now we have considered small area estimation problems only for
continuous-valued response. However, often in practice, response variables
are binary or categorical.\vadjust{\goodbreak} For example, in the SAIPE program, U.S.
Census Bureau is interested in estimating the poverty rates among school
children. The response variable here is binary taking values 1 and 0
depending on whether the child is in poverty or not. More generally, the
response variable may take values in multiple categories. Again, in the
disease mapping context, the response is typically the number of occurrences
of a rare event. Generalized linear models are needed for the analysis of
this kind of data.

Both empirical and hierarchical Bayesian approa\-ches have played an
important role in developing small area estimates for discrete data.
Dempster and Tomberlin (\citeyear{DemTom80}), Farrell, MacGibbon and Tomberlin
(\citeyear{FarMacTom97}) and MacGibbon and
Tomberlin (\citeyear{MacTom89}) have
obtained small area estimates of proportions
based on EB techniques. A general EB formulation for simultaneous estimation
of means from the natural exponential family quadratic variance
function family of distributions is due to Ghosh and Maiti (\citeyear
{GhoMai04}). They
provided also estimated mean squared errors of the small area estimators.
Earlier, for the binary case, Jiang (\citeyear{Jia98}) and Jiang and
Zhang (\citeyear{JiaZha01})
obtained such mean squared error estimators based on the jackknife
approach. On
the other hand, a general hierarchical Bayesian approach based on generalized
linear models in the small area estimation context is due to Ghosh et
al. (\citeyear{Ghoetal98}).

\subsection{Balanced Loss Functions}\label{sec7.3}
HB and EB estimators in the small area context are mostly derived under squared
error loss. As an alternative, Ghosh, Kim and Kim (\citeyear
{GhoKimKim08}) considered the
balanced loss introduced and made popular by Zellner (\citeyear{Zel88,Zel94}). For
simplicity, we go back to the framework of Section \ref{sec2} where we
considered small
area models with equal number of observations within each area. For an
arbitrary estimator $\bT=(T_1,\ldots,T_m)^T$ of $\bth$, the balanced
loss is
given by
$L(\bth,\bT) = m^{-1} [w\Vert \mathbf{y}-\bT\Vert^2 + (1-
w)\Vert\bT- \bth\Vert^2 ]$,
where $\Vert\cdot\Vert$ is the Euclidean norm and $w \in[0, 1]$ is the
known weight.
The choice of
$w$ reflects the relative weight which the experimenter wants to assign
to goodness of
fit and precision of estimation. The extreme cases $w = 0$ and $w = 1$
refer solely to
precision of an estimate and goodness of fit, respectively.

Under the balanced loss with a flat prior for $\bbe$, it follows from Section
\ref{sec2} that the Bayes estimator of $\bth$ is
$\hat{\bth}{}^B_{\operatorname{BAL}}=[1-(1-w)B]\mathbf{y}+(1-w)B\bP
_{\bX}\mathbf{y}$ with
corresponding Bayes risk $m^{-1}E\Vert\hat{\bth}{}^B_{\operatorname{BAL}}{}-\bth\Vert^2=
V[(1-B)+bw^2(m-p)/m]$. An EB estimator is obtained by\vadjust{\goodbreak} substituting the
same estimator $\hat{B}^{\mathit{EB}}=V(m-p-2)/S$ or $(\hat{B}^{\mathit{EB}})^{+}=
\operatorname{min}(\hat{B}^{\mathit{EB}},1)$ of $B$ as given in Section~\ref{sec2},
where we may recall
that $S=\Vert\mathbf{y}-\bP_{\bX}\mathbf{y}\Vert^2$.
The calculation of the Bayes risk
of the resulting EB estimator is similar to that in Section~\ref{sec2}.
The details
are omitted. The special case of the intercept model where $\mathbf{x}_i^T\bbe
=\mu$
for all $i$ was considered in Ghosh, Kim and Kim (\citeyear
{GhoKimKim07,GhoKimKim08}). These authors
also considered constrained Bayes estimators along the lines of Louis
(\citeyear{Lou84})
and Ghosh (\citeyear{Gho92N1}).

\section{Summary and Future Research}\label{sec8}

The paper reviews several normal theory-based small area estimation techniques.
In particular, the role of shrinkage estimation in the small area
context is
highlighted, and different variants of Stein-type shrinkers are discussed.
Both hierarchical and empirical Bayesian methods are presented in the context
of mixed linear models for unbalanced data, and are illustrated with specific
small area problems. Empirical Bayes confidence intervals based on hierarchical
normal models are provided. Extensions of these results to measurement error
models and generalized linear models are also touched upon.

There are several promising areas of future research. As mentioned earlier,
small area estimation needs explicit,
or at least implicit, use of models. These model-based estimates can differ
widely from the direct estimates, especially for areas with very low sample
sizes. One potential drawback of the model-based estimates is that
when aggregated,
the overall estimate for a larger geographical area may be quite different
from the corresponding direct estimate, the latter being usually
believed to be
quite reliable. This is because the original survey was designed to achieve
specified inferential accuracy at this higher level of aggregation. The
problem can become more severe in the event of model failure as often there
is no real check for the validity of the assumed model. Moreover, this
overall agreement with the direct estimates may sometimes be politically
necessary to convince the legislators of the utility of small area estimates.

One way to avoid this problem is the so-called ``benchmarking approach'' which
amounts to modifying these model-based estimates so that one gets the same
aggregate estimate for the larger geographical area. A simple illustration
is to modify the model-based county-level estimates so that one matches the
state-level direct estimate. Currently the most popular approach is the
so-called ``raking'' method which involves multiplying all the small area
estimates by a constant factor so that the weighted total agrees with the
direct estimate. Clearly, this is an ad hoc procedure with very little
statistical foundation.

It appears that constrained Bayes small area estimates (Louis,
\citeyear
{Lou84}; Ghosh,
\citeyear{Gho92N2}) will be particular\-ly appropriate to achieve this
end. Instead of
matching the first two moments from the empirical histogram of Bayes
estimates with those from the posterior histogram of the parameters as in
Louis (\citeyear{Lou84}) or Ghosh (\citeyear{Gho92N1}), one should
require that the aggregate or some
weighted aggregate of these small area estimates should equal the large
area aggregate estimate. This can possibly be achieved even for fairly
complex models. See also Shen and Louis (\citeyear{SheLou98}).

The other interesting issue is to extend the measurement error model much
further so that one can even handle discrete data and also more complex
normal theory models.

\section*{Acknowledgments}
This research was supported in part by NSF\break Grants SES-0631426,
SES-1026165 and SES-0241651. The authors thank the Associate Editor and
a reviewer
for their constructive comments.

%

%

\end{document}